\documentclass{nature}

\newcounter{firstbib}

\usepackage{graphicx}
\usepackage{epsfig}
\usepackage[font={small}]{caption}
\usepackage{tablefootnote}
\usepackage{hyperref}
\usepackage{amsmath}
\usepackage{amssymb}
\usepackage{lineno}

\newcommand{\tej}{\ensuremath{t_{\rm ej}}}

\newcommand{\dr}{\ensuremath{\Delta r}}
\newcommand{\taus}{\ensuremath{\tau_{\rm Sob}}}

\def\Msun{{\rm M}_\odot}

\def\Zsun{{\rm Z}_\odot}

\def\gtorder{\mathrel{\raise.3ex\hbox{$>$}\mkern-14mu
             \lower0.6ex\hbox{$\sim$}}}
\def\ltorder{\mathrel{\raise.3ex\hbox{$<$}\mkern-14mu
             \lower0.6ex\hbox{$\sim$}}}

\def\Angstrom{\textup{\AA}}

\def\hls{iPTF14hls}
\def\snem{SN\,1999em}

\title{Energetic eruptions leading to a peculiar hydrogen-rich explosion of a massive star}

\author{Iair~Arcavi$^{\ref{af:LCO},\ref{af:KITP},\ref{af:UCSB},\ref{af:Einstein}}$,
D.~Andrew~Howell$^{\ref{af:LCO},\ref{af:UCSB}}$,
Daniel~Kasen$^{\ref{af:LBL},\ref{af:UCBPhysics},\ref{af:UCBAstro}}$,
Lars~Bildsten$^{\ref{af:KITP},\ref{af:UCSB}}$,
Griffin~Hosseinzadeh$^{\ref{af:LCO},\ref{af:UCSB}}$,
Curtis~McCully$^{\ref{af:LCO},\ref{af:UCSB}}$,
Zheng~Chuen~Wong$^{\ref{af:LCO},\ref{af:UCSB}}$,
Sarah~Rebekah~Katz$^{\ref{af:LCO},\ref{af:UCSB}}$,
Avishay~Gal-Yam$^{\ref{af:WIS}}$,
Jesper~Sollerman$^{\ref{af:OKC}}$,
Francesco~Taddia$^{\ref{af:OKC}}$,
Giorgos~Leloudas$^{\ref{af:WIS},\ref{af:Dark}}$,
Christoffer~Fremling$^{\ref{af:OKC}}$,
Peter~E.~Nugent$^{\ref{af:LBLcomp},\ref{af:UCBAstro}}$,
Assaf~Horesh$^{\ref{af:HUJI},\ref{af:WIS}}$,
Kunal~Mooley$^{\ref{af:Oxford}}$,
Clare~Rumsey$^{\ref{af:Cambridge}}$,
S.~Bradley~Cenko$^{\ref{af:Goddard},\ref{af:Maryland}}$,
Melissa~L.~Graham$^{\ref{af:Washington},\ref{af:UCBAstro}}$,
Daniel~A.~Perley$^{\ref{af:Dark},\ref{af:Liverpool}}$,
Ehud~Nakar$^{\ref{af:TAU}}$,
Nir~J.~Shaviv$^{\ref{af:HUJI}}$,
Omer~Bromberg$^{\ref{af:TAU}}$,
Ken~J.~Shen$^{\ref{af:UCBAstro}}$,
Eran~O.~Ofek$^{\ref{af:WIS}}$,
Yi~Cao$^{\ref{af:eScience},\ref{af:Washington}}$,
Xiaofeng~Wang$^{\ref{af:Tsinghua}}$,
Fang~Huang$^{\ref{af:Tsinghua}}$,
Liming~Rui$^{\ref{af:Tsinghua}}$,
Tianmeng~Zhang$^{\ref{af:NAOC},\ref{af:CASastro}}$,
Wenxiong~Li$^{\ref{af:Tsinghua}}$,
Zhitong~Li$^{\ref{af:Tsinghua}}$,
Jujia~Zhang$^{\ref{af:Yunnan},\ref{af:CAS}}$,
Stefano~Valenti$^{\ref{af:UCD}}$,
David~Guevel$^{\ref{af:LCO},\ref{af:UCSB}}$,
Benjamin~Shappee$^{\ref{af:Carnegie},\ref{af:HCP}}$,
Christopher~S.~Kochanek$^{\ref{af:OSU},\ref{af:CCAPP}}$,
Thomas~W.-S.~Holoien$^{\ref{af:OSU},\ref{af:CCAPP}}$,
Alexei~V. Filippenko$^{\ref{af:UCBAstro},\ref{af:Miller}}$,
Rob~Fender$^{\ref{af:Oxford}}$,
Anders~Nyholm$^{\ref{af:OKC}}$,
Ofer~Yaron$^{\ref{af:WIS}}$,
Mansi~M.~Kasliwal$^{\ref{af:CIT}}$,
Mark~Sullivan$^{\ref{af:Southampton}}$,
Nadja~Blagorodnova$^{\ref{af:CIT}}$,
Richard~S.~Walters$^{\ref{af:CIT}}$,
Ragnhild~Lunnan$^{\ref{af:CIT}}$,
Danny~Khazov$^{\ref{af:WIS}}$,
Igor~Andreoni$^{\ref{af:Swinburne},\ref{af:CAASTRO},\ref{af:AAO}}$,
Russ~R.~Laher$^{\ref{af:IPAC}}$,
Nick~Konidaris$^{\ref{af:Carnegie}}$,
Przemek~Wozniak$^{\ref{af:LANL}}$,
and
Brian~Bue$^{\ref{af:JPL}}$
}

\begin{document}

\maketitle

\begin{affiliations}
 \item Las Cumbres Observatory, Goleta, CA 93117, USA\label{af:LCO}
 \item Kavli Institute for Theoretical Physics, University of California, Santa Barbara, CA 93106, USA\label{af:KITP}
 \item Department of Physics, University of California, Santa Barbara, CA 93106, USA\label{af:UCSB}
 \item Einstein Fellow\label{af:Einstein}
 \item Nuclear Science Division, Lawrence Berkeley National Laboratory, Berkeley, CA 94720, USA\label{af:LBL}
 \item Department of Physics, University of California, Berkeley, CA 94720, USA\label{af:UCBPhysics}
 \item Department of Astronomy, University of California, Berkeley, CA 94720-3411, USA\label{af:UCBAstro}
 \item Department of Particle Physics and Astrophysics, The Weizmann Institute of Science, Rehovot, 76100, Israel\label{af:WIS}
 \item The Oskar Klein Centre, Department of Astronomy, Stockholm University, AlbaNova, SE-10691 Stockholm, Sweden\label{af:OKC}
 \item Dark Cosmology Centre, Niels Bohr Institute, University of Copenhagen, Juliane Maries vej 30, 2100 Copenhagen, Denmark\label{af:Dark}
 \item Computational Research Division, Lawrence Berkeley National Laboratory, Berkeley, CA 94720, USA\label{af:LBLcomp}
 \item Racah Institute of Physics, The Hebrew University of Jerusalem, Jerusalem 91904, Israel\label{af:HUJI}
 \item Department of Physics, Astrophysics, University of Oxford, Denys Wilkinson Building, Oxford, OX1 3RH, UK\label{af:Oxford}
 \item Astrophysics Group, Cavendish Laboratory, 19 J. J. Thomson Avenue, Cambridge, CB3 0HE, UK\label{af:Cambridge}
 \item Astrophysics Science Division, NASA Goddard Space Flight Center, Code 661, Greenbelt, MD 20771, USA\label{af:Goddard}
 \item Joint Space-Science Institute, University of Maryland, College Park, MD 20742, USA\label{af:Maryland}
 \item Department of Astronomy, University of Washington, Box 351580, U.W., Seattle, WA 98195-1580, USA\label{af:Washington}
 \item Astrophysics Research Institute, Liverpool John Moores University, IC2, Liverpool Science Park, 146 Brownlow Hill, Liverpool L3 5RF, UK\label{af:Liverpool}
 \item The Raymond and Beverly Sackler School of Physics and Astronomy, Tel Aviv University, Tel Aviv 69978, Israel\label{af:TAU}
 \item eScience Institute, University of Washington, Box 351570, U.W., Seattle, WA 98195-1580, USA\label{af:eScience}
 \item Physics Department and Tsinghua Center for Astrophysics (THCA), Tsinghua University, Beijing, 100084, China\label{af:Tsinghua}
 \item Key Laboratory of Optical Astronomy, National Astronomical Observatories of China, Chinese Academy of Sciences, Beijing 100012, Chin\label{af:NAOC}
 \item School of Astronomy and Space Science, University of Chinese Academy of Sciences, Beijing 101408, China\label{CASastro}
 \item Yunnan Observatories, Chinese Academy of Sciences, Kunming 650011, China\label{af:Yunnan}
 \item Key Laboratory for the Structure and Evolution of Celestial Objects, Chinese Academy of Sciences, Phoenix Mountain, East District, Kunming, Yunnan 650216, China\label{af:CAS}
 \item Department of Physics, University of California, 1 Shields Ave, Davis, CA 95616, USA\label{af:UCD}
 \item Carnegie Observatories, 813 Santa Barbara Street, Pasadena, CA 91101, USA\label{af:Carnegie}
 \item Hubble Fellow, Carnegie-Princeton Fellow\label{af:HCP}
 \item Department of Astronomy, The Ohio State University, 140 West 18th Avenue, Columbus, OH 43210, USA\label{af:OSU}
 \item Center for Cosmology and AstroParticle Physics (CCAPP), The Ohio State University, 191 W. Woodruff Ave., Columbus, OH 43210, USA\label{af:CCAPP}
 \item Miller Senior Fellow, Miller Institute for Basic Research in Science, University of California, Berkeley, CA 94720, USA\label{af:Miller}
 \item Cahill Center for Astrophysics, California Institute of Technology, Pasadena, CA 91125, USA\label{af:CIT}
 \item Department of Physics and Astronomy, University of Southampton, Southampton SO17 1BJ, UK\label{af:Southampton}
 \item Centre for Astrophysics and Supercomputing, Swinburne University of Technology, PO Box 218, VIC 3122, Australia\label{af:Swinburne}
 \item ARC Centre of Excellence for All-sky Astrophysics (CAASTRO)\label{af:CAASTRO}
 \item Australian Astronomical Observatory, PO Box 915, North Ryde, NSW 1670, Australia\label{af:AAO}
 \item Spitzer Science Center, California Institute of Technology, MS 314-6, Pasadena, CA 91125, USA\label{af:IPAC}
 \item Space and Atmospheric Sciences Group, Mail Stop D466, Los Alamos National Laboratory, Los Alamos, NM 87545, USA\label{af:LANL}
 \item Jet Propulsion Laboratory, California Institute of Technology, Pasadena, CA 91109, USA\label{af:JPL}
\end{affiliations}



\clearpage

\begin{abstract}
Every supernova hitherto observed has been considered to be the terminal explosion of a star. Moreover, all supernovae with absorption lines in their spectra show those lines decreasing in velocity over time, as the ejecta expand and thin, revealing slower moving material that was previously hidden. In addition, every supernova that exhibits the absorption lines of hydrogen has one main light-curve peak, or a plateau in luminosity, lasting approximately 100 days before declining\cite{Arcavi2016}. Here we report observations of {\hls}, an event that has spectra identical to a hydrogen-rich core-collapse supernova, but characteristics that differ extensively from those of known supernovae. The light curve has at least five peaks and remains bright for more than 600 days; the absorption lines show little to no decrease in velocity; and the radius of the line-forming region is more than an order of magnitude bigger than the radius of the photosphere derived from the continuum emission. These characteristics are consistent with a shell of several tens of solar masses ejected by the star at supernova-level energies a few hundred days before a terminal explosion. Another possible eruption was recorded at the same position in 1954. Multiple energetic pre-supernova eruptions are expected to occur in stars of 95--130 solar masses, which experience the pulsational pair instability\cite{Barkat1967,Heger2002,Woosley2007,Woosley2017}. That model, however, does not account for the continued presence of hydrogen, or the energetics observed here. Another mechanism for the violent ejection of mass in massive stars may be required.
\end{abstract}


On 2014 September 22.53 UT (universal time used throughout), the intermediate Palomar Transient Factory (iPTF) wide-field camera survey\cite{Law2009,Rau2009} discovered {\hls} at right ascension $\alpha_{\rm J2000}=09$h~$20$m~$34.30$s and declination $\delta_{\rm J2000}=+50^{\circ}41'46.8''$, at an $R$-band magnitude of $17.716 \pm 0.033$ (Extended Data Fig. \ref{EDfig:discovery}). We have no observations of this position between 2014 May 28 and September 22, inducing an approximately 100-day uncertainty in the explosion time, so we use the discovery date as a reference epoch for all phases. We adopt a redshift of $z=0.0344$, determined from narrow host-galaxy features, corresponding to a luminosity distance\cite{Planck2015} of 156\,Mpc. 

On 2015 January 8, {\hls} was classified as a supernova of type II-P, on the basis of prominent, broad, Balmer series P-Cygni lines in an optical spectrum\cite{Li2015}. So far, type~II-P supernovae have been the only events ever observed to produce such spectra. In a type~II-P supernova, the core of a massive star collapses to create a neutron star, sending a shock wave through the outer hydrogen-rich envelope and ejecting it. The shock ionizes the ejecta, which later expand, cool and recombine. The photosphere follows the recombination front, which is at a roughly constant temperature ($T\approx6,000$\,K) as it makes its way inward in mass through the expanding ejecta\cite{Popov1993} (that is, the photosphere is moving from material that is further out from the exploding star towards material that is further in, but the material inside the photosphere is expanding in the meantime). This leads to the approximately 100-day `plateau' phase of roughly constant luminosity in the light curve and prominent hydrogen P-Cygni features in the spectrum. 

\begin{figure}
\centering
\includegraphics[width=150mm]{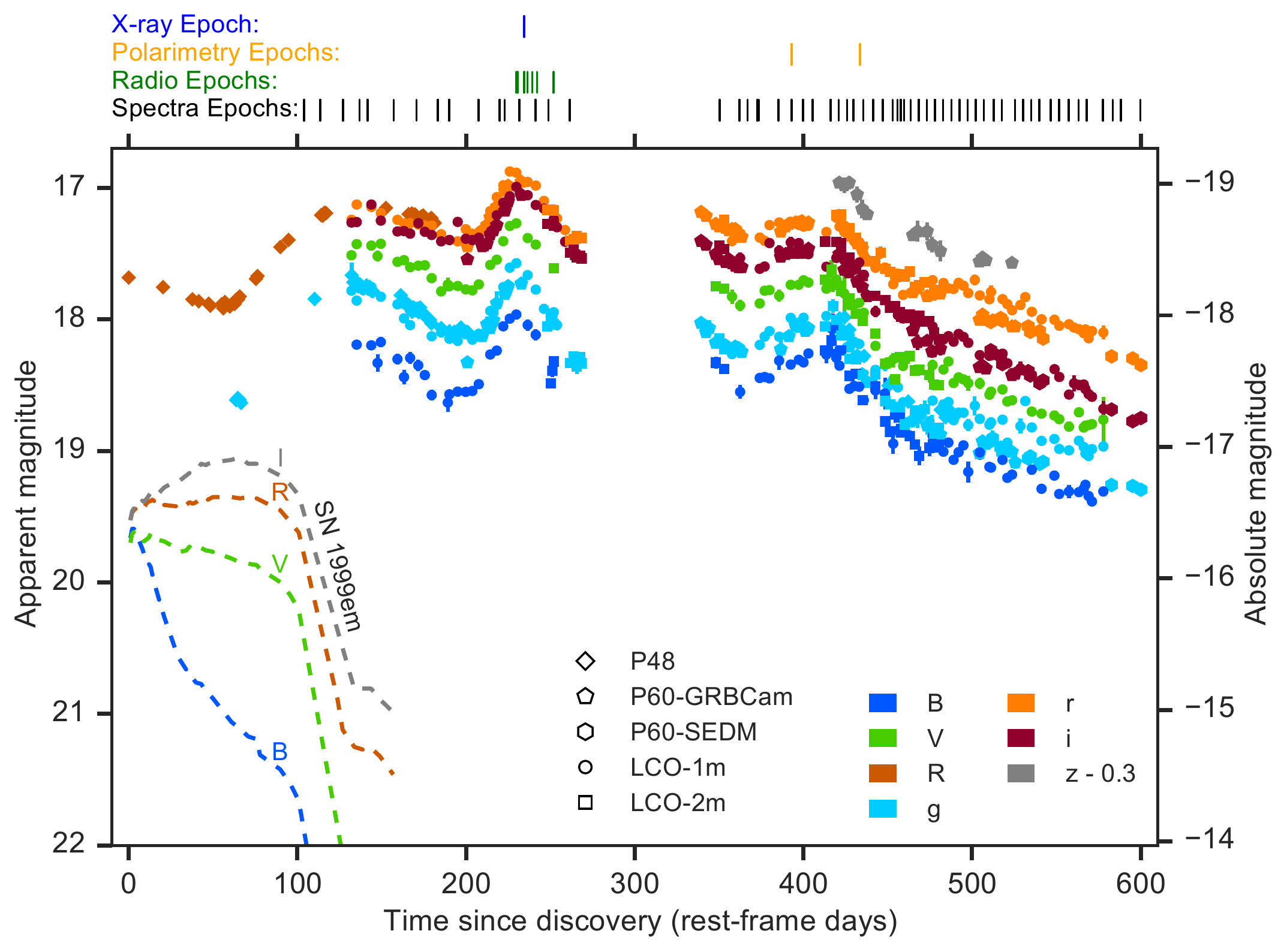}
\caption{{\bf Light curves of {\hls}.} The type~II-P supernova {\snem} is shown in dashed lines\cite{Leonard2002}, according to the ordinate axis at right. Data from the same day, instrument, and filter are averaged for clarity. The Spectral Energy Distribution Machine (SEDM) $i$-band data are shifted by $+0.3$ mag to compensate for filter differences. Unlike any known supernova, {\hls} has at least five distinct peaks in its light curve (at approximately 140 days, 220 days, and 410 days after discovery, before discovery as indicated by the $R$-band light curve, and while it was behind the Sun between day 260 and day 340 after discovery). Error bars denote $1\sigma$ uncertainties. 
\label{fig:photometry}
}
\end{figure}

{\hls}, although identical to type~II-P supernovae in its spectroscopic features, has several properties never before seen in a supernova. Instead of a 100-day plateau, the light curve of {\hls} lasts over 600 days and has at least five distinct peaks during which the luminosity varies by as much as $50\%$ (Fig. \ref{fig:photometry}). Blackbody fits to the broad-band optical {\it BVgi} photometry of {\hls} (see Methods) indicate a roughly constant effective temperature of $5,000$--$6,000$\,K, the same as the hydrogen-recombination temperature typically seen in type~II-P supernovae. However, the inferred bolometric luminosity of a few times $10^{42}$\,erg\,s$^{-1}$ is on the high end of the range for typical type~II-P supernovae\cite{Bersten2009}, and the total radiated energy of $2.20_{-0.05}^{+0.03}\times10^{50}$\,erg emitted during the $450$ days of our multi-band optical coverage is a few times larger than that of any known type~II-P supernova. Given the uncertainty in explosion time of {\hls}, the discrepancies with type II-P supernova timescales and energetics may be even larger.

\begin{figure}
\centering
\includegraphics[width=183mm]{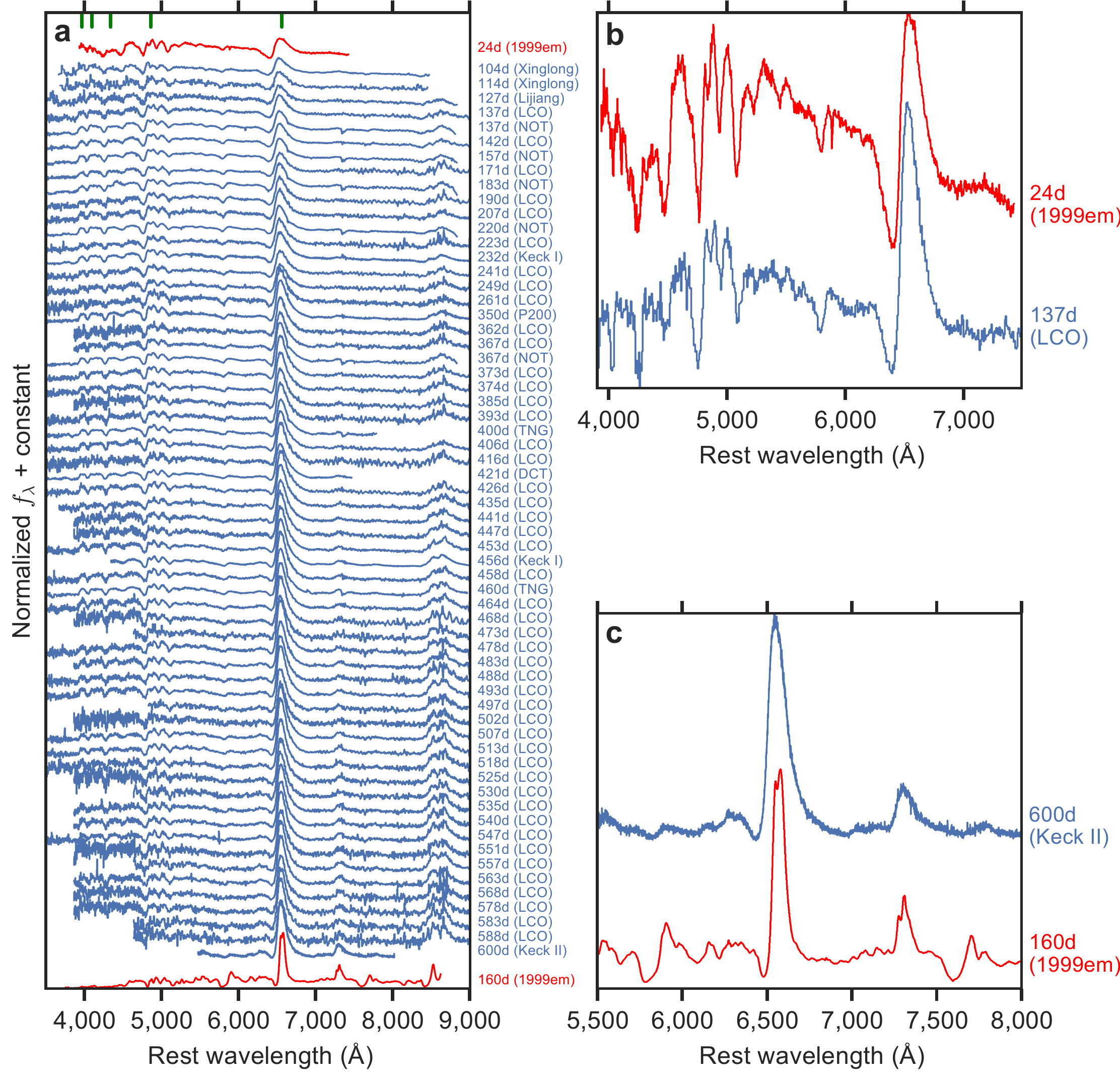}
\caption{{\bf Spectroscopic sequence of {\hls}.} Our full sequence (\textbf{a}; blue) is shown with select spectra highlighted (\textbf{b},\textbf{c}), in terms of normalized flux density as a function of rest-frame wavelength. The spectra are binned in wavelength and shifted in flux density for clarity. Phases are noted in rest-frame days since discovery, with the telescopes used in parentheses. Spectra of the type~II-P supernova {\snem}\cite{Leonard2002} (red) are shown with phases noted in rest-frame days since explosion. Balmer series hydrogen-line wavelengths are denoted in green (\textbf{a}). {\hls} is very similar spectroscopically to a type~II-P supernova but evolves much more slowly. The spectral evolution is very smooth (\textbf{a}).
\label{fig:spectra}
}
\end{figure}

\begin{figure}
\centering
\includegraphics[width=150mm]{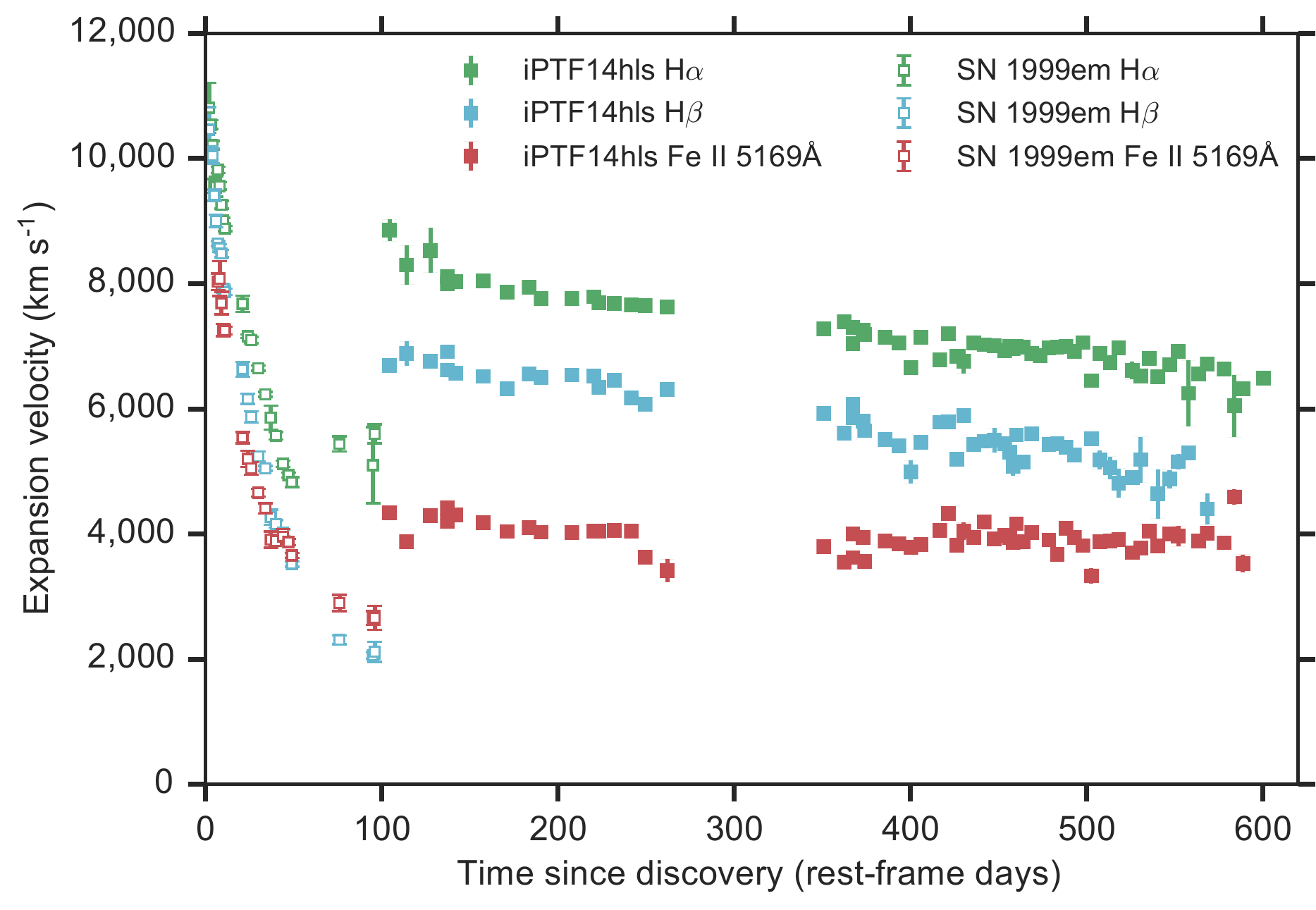}
\caption{{\bf Expansion velocities as a function of time.} Velocities are measured from the P-Cygni absorption component of three different spectral lines (see Methods) for {\hls} (filled symbols) and the prototypical type~II-P supernova \snem\cite{Leonard2002} (empty symbols). Error bars denote $1\sigma$ uncertainties for $n=1,000$ samplings of the endpoints and are sometimes smaller than the marker size. The velocities seen for {\hls} evolve much more slowly compared with \snem.
\label{fig:velocities}
}
\end{figure}

The spectroscopic evolution of {\hls} is even harder to understand. It is a factor of approximately $10$ slower than that of type~II-P supernovae (Fig. \ref{fig:spectra}); for example, the spectrum of {\hls} at 600 days looks like that of a normal type~II-P supernova at 60 days (Extended Data Fig. \ref{EDfig:superfits}). In all previously observed supernovae, the faster material is outside --- spectra show a decrease of all measured velocities with time (by a factor of  approximately $3$ over $100$ days) as the material expands and thins, and the photosphere moves inward in mass revealing deeper, slower-moving material. In {\hls}, velocities of hydrogen decline by only 25\%, from $8,000$\,km\,s$^{-1}$ 
to $6,000$\,km s$^{-1}$ over $600$ days, while the iron lines stay at a constant velocity of $4,000$\,km\,s$^{-1}$ (Fig. \ref{fig:velocities}).  

It is usual to see hydrogen lines at higher velocity than iron lines owing to optical depth effects. But eventually, as the material expands and thins, hydrogen should be seen at lower velocity where the iron was previously seen (Extended Data Fig. \ref{EDfig:vgradient}). If the ejecta are expanding in size by a factor of approximately $6$ from day $100$ to day $600$, in the absence of an additional energy source, an inward-moving photosphere scanning through the ejecta in velocity must occur.  

An observation of constant velocity in a supernova can thus be caused by: (1) a central engine pushing material from the inside, sweeping the ejecta into a thin, dense shell\cite{Kasen2010magnetar,Dexter2013}, or (2) the spectral lines may correspond to ejecta that are far above the photosphere and detached from it. One-dimensional central-engine models compress the iron and hydrogen lines to the same velocity, which is not the case for {\hls} (though multi-dimensional effects could alter this prediction). The spectral line evolution can more readily be explained if the lines are formed by ejecta from a prior eruption (that happened a few years before the discovery of {\hls}) that are detached from the continuum-emitting photosphere, which was formed in the terminal explosion (see Methods).

\begin{figure}
\centering
\includegraphics[width=150mm]{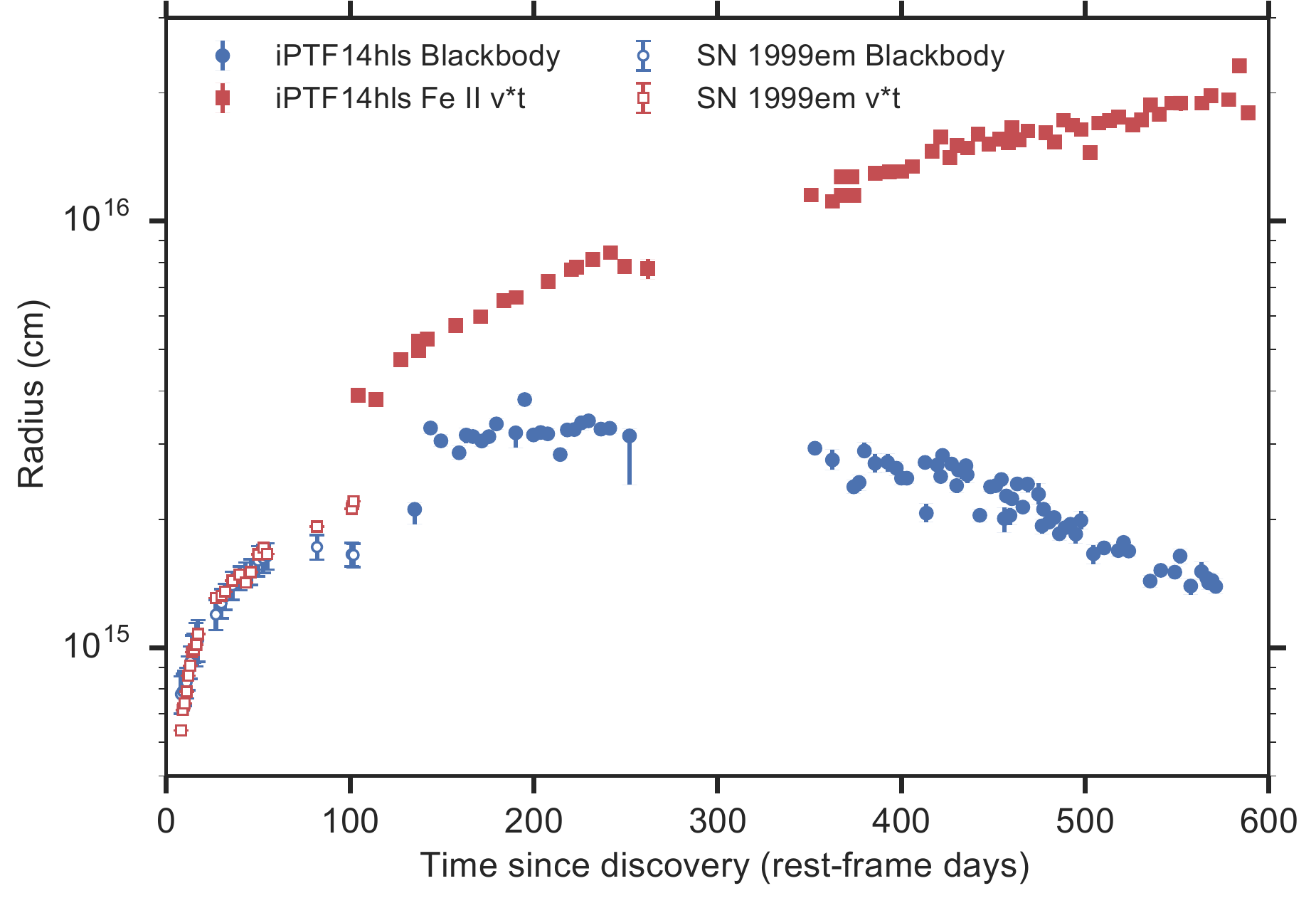}
\caption{{\bf The photospheric radius of {\hls}.} The radius is estimated in two different ways: (1) using blackbody fits to the broad-band {\it BVgi} photometry (blue) and (2) using the derived expansion velocities of Fe~II $\lambda=5,169\Angstrom$ times the elapsed rest-frame time ($vt$) since discovery (red). The same quantities are shown for the type~II-P supernova {\snem} (empty symbols; after correcting for the blackbody dilution factor)\cite{Leonard2002}. Error bars denote $1\sigma$ uncertainties and are sometimes smaller than the marker size. For {\snem} the radii overlap as expected, but for {\hls} they diverge, indicating that the line-forming region may be detached from the photosphere.
\label{fig:radius}
}
\end{figure}

We estimate the position of the line-forming region as $vt$, where $v$ is the observed expansion velocity of the material at time $t$. For type~II-P supernovae, this radius, when using the iron line velocities, is the same as the photospheric radius obtained by blackbody fits to the continuum emission, up to an order-unity ``blackbody dilution factor''\cite{Kirshner1975,Eastman1996,Dessart2005}. For {\hls}, the $vt$-inferred radius is instead larger than the blackbody-inferred radius by an order of magnitude on day $600$ (Fig. \ref{fig:radius}). The fact that the two radii are so different from each other indicates that the line-forming region in {\hls} is indeed spatially detached from the continuum-emitting photosphere, in contrast to what is observed in all known type~II-P supernovae.  

The observations are thus consistent with the line-forming material being ejected in a massive and very energetic pre-supernova outburst, specifically in a shell on the order of a few tens of solar masses (see Methods). However, this requires a kinetic energy of about $10^{52}$ erg, normally associated with (or even exceeding that of) a supernova. Further evidence for a third, even earlier, explosion comes from an outburst of $R$-band magnitude $M_R\approx-15.6$ detected at the position of {\hls} in 1954 (formally a 2.2$\sigma$ detection, though this is probably an underestimate owing to photographic nonlinearity; see Methods). 

Another question is what is powering the light curve of {\hls}. Strong asymmetry may induce a luminosity increase in a particular direction. However, we do not detect any substantial polarization that would be indicative of asymmetry in the explosion (see Methods). Another energy source in {\hls} compared to normal type~II-P supernovae could come from the interaction of the ejecta with previously ejected shells. However, in cases of supernovae interacting with dense circumstellar material, the interaction dominates the spectra in the form of a strong continuum together with broad, intermediate and narrow components of the Balmer emission lines\cite{Schlegel1990,Kiewe2012}. None of these features is seen in the spectra of {\hls} (Fig. \ref{fig:spectra}; Extended Data Fig. \ref{EDfig:deimos}). We find no evidence of X-ray or radio emission (which are possible additional indicators of strong interaction)\cite{Chevalier2006} in observations taken during the brightest peak of the optical light curve (see Methods). It is possible that any signs of interaction are being reprocessed by overlying, previously ejected material.  

Either way, the progenitor of {\hls} likely experienced multiple energetic eruptions over the last decades of its life. Energetic eruptions are expected in stars with initial masses of about $95$--$130$ solar masses, which undergo an instability arising from the production of electron-positron pairs\cite{Barkat1967}. Interaction between the different shells and/or the supernova ejecta and the shells can produce a variety of luminous long-lived transients with highly structured light curves\cite{Woosley2007,Woosley2017} similar to that of {\hls}. Such pulsational-pair instability supernovae are expected to occur in low-metallicity environments; indeed, {\hls} occurred on the outskirts of a low-mass star-forming galaxy, possibly of low metal content (see Methods).

However, models of stars undergoing the pulsational pair instability eject most of the hydrogen envelope in the first eruption\cite{Woosley2017}, whereas for {\hls} a few tens of solar masses of hydrogen were retained in the envelope after the 1954 outburst. Another problem is that pulsational pair instability models can account for up to about $4\times10^{51}$\,erg of kinetic energy in all eruptions together, while about $10^{52}$\,erg are required just for the most recent eruption that ejected the line-forming region of {\hls} (see Methods). 

{\hls} demonstrates that stars in the local Universe can undergo very massive eruptions in the decades leading to their collapse yet, surprisingly, maintain a massive hydrogen-rich envelope for most of this period. Current models of massive star evolution and explosion need to be modified, or a completely new picture needs to be put forward, to account for the energetics of {\hls}, its lack of strong interaction signatures and the inferred amount of hydrogen it retained toward the end of its life.

%
%

%



 
\begin{addendum}


\item[Author Contributions]
I.~Arcavi initiated the study, triggered follow-up observations, reduced data, performed the analysis and wrote the manuscript. 
D.A.H. is PI of the LCO Supernova Key Project through which all of the LCO data were obtained; he also assisted with interpretation and the manuscript. 
D.~Kasen and L.B. assisted with theoretical models, data interpretation, and with the manuscript. 
G.H. and C.M. assisted with obtaining and reducing LCO data.
Z.C.W. first flagged the supernova as interesting.
S.R.K. performed the spectral expansion velocity measurements.
A.G.-Y. is the PI for core-collapse supernovae in iPTF and assisted with interpretation.
J.S. and F.T. obtained the NOT spectra and polarimetry data and assisted with the manuscript.
G.L. reduced the polarimetry data.
C.F. reduced the P60 data.
P.E.N. discovered the 1954 eruption image of {\hls}, helped obtain the host-galaxy spectrum, and is a Co-PI of the Keck proposal under which it and one of the supernova spectra were obtained.
A.H. obtained and reduced the VLA data and is PI of the program through which the data were obtained.
K.M. and C.R. obtained and reduced the AMI data.
S.B.C. obtained and reduced the XRT data.
M.L.G. obtained and reduced Keck spectra.
D.A.P. performed the host-galaxy analysis and assisted with the manuscript.
E.N., O.B., N.J.S., and K.J.S. assisted with theoretical interpretation and the manuscript.
E.O.O. helped with interpretation and the manuscript.
Y.C. built the real-time iPTF image-subtraction pipeline and obtained P200 observations.
X.W., F.H., L.R., T.Z., W.L., Z.L., and J.Z. obtained and reduced the Xinglong, Lijiang, and TNT data.
S.V. built the LCO photometric and spectroscopic reduction pipelines and assisted with LCO observations, interpretation, and the manuscript.
D.G. assisted with the POSS image analysis.
B.S., C.S.K., and T.W.-S.H. obtained and reduced the ASAS-SN pre-discovery limits.
A.V.F. is a Co-PI of the Keck proposal under which the host-galaxy spectrum and one of the supernova spectra were obtained; he also helped with the manuscript.
R.F. is PI of the program through which the AMI data were obtained.
A.N. helped scan for iPTF candidates and assisted with the manuscript.
O.Y. is in charge of the iPTF candidate scanning effort.
M.M.K. led the work for building iPTF.
M.S. wrote the pipeline used to reduce P48 data.
N.B. and R.S.W. obtained P60 SEDM photometry.
R.N., D.~Khazov, and I.~Andreoni obtained P200 observations.
R.R.L. contributed to building the P48 image-processing pipeline.
N.K. was a main builder of the P60 SEDM.
P.W. and B.B. helped build the machine-learning algorithms that identify iPTF supernova candidates.
 
\item[Author Information] Reprints and permissions information is available at www.nature.com/reprints. The authors declare that they have no competing financial interests. Correspondence and requests for materials should be addressed to I.~Arcavi (arcavi@gmail.com).

\end{addendum}


\clearpage

\newpage

\newpage

\newpage

\clearpage

\begin{methods}

\subsection{Discovery.}
\label{sec:disc}

The intermediate Palomar Transient Factory (iPTF) first detected {\hls} on 2014 Sep. 22.53 (Extended Data Fig. \ref{EDfig:discovery}) using the iPTF real-time image-subtraction pipeline\cite{Cao2016}. No source was seen at that position when it was previously visited by iPTF and by the All Sky Automated Survey for Supernova (ASAS-SN)\cite{Shappee2014} on 2014 May 6.19 and 2014 May 20--28 down to $3\sigma$ limiting magnitudes of $R < 20.95$ and $V < 18.7$, respectively. The source was observed by iPTF again on 2014 Oct. 13, Oct. 31, Nov. 4, and Nov. 10 before being saved and given a name as part of routine iPTF transient scanning. On 2014 Nov. 18, {\hls} was independently discovered by the Catalina Real-Time Transient Survey\cite{Drake2009} as CSS141118:092034+504148, and later the event was reported to the Transient Name Server as AT\,2016bse and Gaia16aog. On 2015 Feb. 3, upon routine LCO rescanning of previously saved iPTF candidates, we noticed the peculiar decline and subsequent rise of the light curve, and began an extensive campaign of spectroscopic and multi-band photometric follow-up observations.

\subsection{Follow-up imaging.}
\label{sec:imaging}

Follow-up imaging was obtained with the Palomar 48-inch Oschin Schmidt telescope (P48), the Palomar 60-inch telescope (P60)\cite{Cenko2006} using both the GRBCam and the SED Machine (SEDM) instruments, the Las Cumbres Observatory (LCO)\cite{Brown2013} network 1-m and 2-m telescopes, and the 0.8-m Tsinghua University-NAOC telescope (TNT)\cite{Huang2012} at the Xinglong Observatory. The TNT photometry is presented (together with CSS and Gaia photometry downloaded from their respective websites) in Extended Data Figure \ref{EDfig:addedphot}. P48 images were first pre-processed by the Infrared Processing and Analysis Center (IPAC)\cite{Laher2014}. Image subtraction and point-spread-function (PSF) fitting were then performed\cite{Sullivan2006} using pre-explosion images as templates. Magnitudes were calibrated to observations of the same field by the Sloan Digital Sky Survey (SDSS) DR10\cite{Ahn2014}. P60 images were pre-processed using a PyRAF-based pipeline\cite{Cenko2006}. Image subtraction, photometry extraction, and calibration were performed with the {\tt FPipe} pipeline\cite{Fremling2016} using SDSS images as references. LCO images were pre-processed using the Observatory Reduction and Acquisition Control Data Reduction pipeline (ORAC-DR)\cite{Jenness2015} up to 2016 May 4, and using the custom Python-based {\tt BANZAI} pipeline afterward. Photometry was then extracted using the PyRAF-based {\tt LCOGTSNpipe} pipeline\cite{Valenti2016} to perform PSF fitting and calibration to the AAVSO Photometric All-Sky Survey\cite{Henden2009} for $BV$-band data and SDSS DR8\cite{Aihara2011} for $gri$-band data. TNT images were reduced with standard IRAF routines; PSF fitting was performed using the {\tt SNOoPy} package and calibrated to the SDSS DR9\cite{Ahn2012} transformed to the Johnson system\cite{Chonis2008}. We correct all photometry for Milky Way extinction\cite{Schlafly2011} extracted via the NASA Extragalactic Database (NED). Pre-explosion nondetection limits are presented in Extended Data Figure \ref{EDfig:limits}.

{\noindent}We fit a blackbody spectral energy distribution (SED) to every epoch of LCO photometry containing at least three of the {\it BVgi} filters obtained within $0.4$ days of each other (we exclude $r$-band and $R$-band data from the fits owing to contamination from the H$\alpha$ line). For each epoch we perform a blackbody fit using Markov Chain Monte Carlo simulations through the Python {\tt emcee} package\cite{Foreman-Mackey2012} to estimate the blackbody temperature and radius at the measured distance to {\hls} of 156\,Mpc. 

\subsection{Follow-up spectroscopy.}
\label{sec:spec}

Spectra of {\hls} were obtained with the Floyds instrument mounted on the northern LCO 2-m telescope\cite{Brown2013}, the Andalucia Faint Object Spectrograph and Camera (ALFOSC) mounted on the 2.5-m Nordic Optical Telescope (NOT), the Device Optimized for the LOw RESolution (DOLoRes) mounted on the 3.6-m Telescopio Nazionale Galileo (TNG), the Low Resolution Imaging Spectrometer (LRIS)\cite{Oke1995} mounted on the Keck-I 10-m telescope, the DEep Imaging Multi-Object Spectrograph (DEIMOS)\cite{Faber2003} mounted on the Keck-II 10-m telescope, the Double Beam Spectrograph (DBSP)\cite{Oke1982} mounted on the Palomar 200-inch telescope (P200), the Beijing Faint Object Spectrograph and Camera (BFOSC) on the Xinglong 2.16-m telescope of the National Astronomical Observatories of China, the Yunnan Faint Object Spectrograph and Camera (YFOSC) on the Lijiang 2.4-m telescope of the Yunnan Observatories, and the DeVeny spectrograph mounted on the 4.3-m Discovery Channel Telescope (DCT). The Floyds spectra were reduced using the PyRAF-based {\tt floydsspec} pipeline. The ALFOSC and DOLORES spectra were reduced using custom MATLAB pipelines. The LRIS spectra were reduced using the IDL {\tt LPipe} pipeline. The DEIMOS spectrum was reduced using a modified version of the {\tt DEEP2} pipeline\cite{cooper2012,newman2012} combined with standard PyRAF and IDL routines for trace extraction, flux calibration and telluric correction. The DBSP spectrum was reduced using custom IRAF and IDL routines. The BFOSC, YFOSC and DeVeny spectra were reduced using standard IRAF procedures. No Na~I D absorption is seen at the redshift of the host galaxy, indicating very low host-galaxy extinction at the supernova position.

{\noindent}We fit each {\hls} spectrum to a library of Type~II supernovae (which includes a full set of {\snem} spectra\cite{Leonard2002}) using Superfit\cite{Howell2005}. We then calculate the average best-fit supernova phase, weighing all the possible fits by their corresponding fit scores. We repeat this process for cutouts of the {\hls} spectra centered around the H$\alpha$, H$\beta$, and Fe II $\lambda5169$ features (separately). The weighted-average best-fit phases for each cutout are presented in Extended Data Figure \ref{EDfig:superfits}. {\hls} can be seen to evolve more slowly than other Type II supernovae by a factor of $\approx$10 when considering the entire spectrum, as well as when considering the H$\beta$ and the Fe~II 5169\,$\Angstrom$ features separately, and by a factor of 6--7 when considering the H$\alpha$ emission feature separately.

{\noindent}Expansion velocities for different elements in {\hls} were measured by fitting a parabola around the minimum of the absorption feature of their respective P-Cygni profiles. The difference between the minimum of the best-fit parabola and the rest wavelength of the line was translated to an expansion velocity. The endpoints of each parabolic fit were chosen manually for each line, so that they would remain the same for all spectra. Uncertainties in the velocities were estimated by randomly varying these endpoints 1000 times by $\pm5\,\Angstrom$ around their original values.

\subsection{{\hls} is probably not powered by interaction.}
\label{sec:xraysandradio}

As mentioned in the main text, interaction between supernova ejecta and pre-existing, dense circumstellar material (CSM) could cause an increase in luminosity. However, {\hls} does not display the spectral line profiles typically seen in such cases (Extended Data Fig. \ref{EDfig:deimos}).

{\noindent}In some interaction models the collision of the supernova ejecta and the CSM occurs outside the broad-line forming region, diluting the line emission. Focusing on the $\approx50\%$ luminosity increase of {\hls} between rest-frame day 207 and 232 after discovery (Fig. \ref{fig:photometry}), we find that the spectra taken on day 207 and day 232 are identical up to a global normalization factor. This indicates that the increase in luminosity is equal at all wavelengths, in contrast to the expected line dilution from interaction (Extended Data Fig. \ref{EDfig:addedflux}).

{\noindent}Additional possible indicators of interaction are strong X-ray and/or radio emission. We observed the location of {\hls} with the X-Ray Telescope (XRT)\cite{Burrows2005} onboard the \textit{Swift} satellite\cite{Gehrels2004} on 2015 May 23.05. A total $4.9$\,ks of live exposure time was obtained on the source. We use online analysis tools\cite{Evans2007,Evans2009} to search for X-ray emission at the location of {\hls}. No source is detected with an upper limit on the 0.3--10.0\,keV count rate of $< 2.3 \times10^{-3}$\,ct\,s$^{-1}$.  Assuming a power-law spectrum with a photon index of $\Gamma\,=\,2$ and a Galactic H column density\cite{Willingale2013} of $1.4 \times 10^{20}$\,cm$^{-2}$, this corresponds to an upper limit on the unabsorbed 0.3--10.0\,keV flux of $f_{X}\,<\,8.4 \times 10^{-14}$\,erg\,cm$^{-2}$\,s$^{-1}$. At the luminosity distance of {\hls} this corresponds to a luminosity limit of $L_{X}\,<\,2.5 \times 10^{41}$\,erg\,s$^{-1}$ (which is roughly $10^{-2}$ of the peak bolometric luminosity). The lack of X-ray emission disfavors strong interaction in {\hls}, though some interacting supernovae display X-ray emission fainter than the limit we deduce here\cite{Margutti2016}.

{\noindent}We observed {\hls} also with the Arcminute Microkelvin Imager Large Array (AMI-LA)\cite{Zwart2008} at 15 GHz on 2015 May 18.59, May 19.77, May 23.63, May 25.65, May 28.66, and May 31.62. 3C48 and J2035+1056 were used as the flux/bandpass and phase calibrators, respectively. RFI excision and calibration of the raw data were done with a fully automated pipeline {\tt AMI-REDUCE}\cite{Davies2009, Perrott2013}. The calibrated data for the supernova were imported into CASA and imaged independently for each epoch into $512 \times 512$ pixel maps ($4''$ per pixel) using the {\tt clean} task. A similar imaging scheme was used for the concatenated data from all the epochs as well. The supernova was not detected on any of the individual epochs, with 3$\sigma$ upper limits of 60--120 $\mu$Jy. The combined 3$\sigma$ upper limit is 36 $\mu$Jy. There is a 5--10\% absolute flux calibration uncertainty that we have not considered in these upper limits. On 2016 June 10, {\hls} was observed with the VLA at $6.1$\,GHz. The VLA data were reduced using standard CASA software routines where J0920+4441 and 3C286 were used as phase and flux calibrators. No radio emission was observed at the supernova position to a $3\sigma$ upper limit of $21.3\,\mu$Jy. At the luminosity distance of {\hls}, this corresponds to $6.2\times10^{26}$\,erg\,s$^{-1}$\,Hz$^{-1}$, which is fainter than the radio emission of most interacting supernovae\cite{Margutti2016}.

{\noindent}We conclude that {\hls} does not show any of the signatures normally seen in supernovae powered by interaction.

\subsection{A possible central-engine power source for {\hls}.}
\label{sec:centralengine}

A central engine such as the spindown of a magnetar\cite{Ostriker1969,Kasen2010magnetar,Woosley2010} or fallback accretion onto a black hole\cite{Colgate1971,Dexter2013} created after core collapse (assuming the material falling back has sufficient angular momentum to form a disk) could inject power to the supernova, although (as noted in the main text) this may fail to reproduce the observed iron and hydrogen line velocity difference. A magnetar (with an initial spin period of ${\approx}5$--$10$\,ms and a magnetic field of ${\approx}(0.5$--$1)\times10^{14}$\,Gauss) can produce the observed average luminosity and timescale of {\hls}\cite{Kasen2010magnetar}. However, the analytical magnetar light curve required to fit the late-time decline overpredicts the early-time emission of {\hls} (Extended Data Fig. \ref{EDfig:addedphot}) and produces a smooth rather than variable light curve\cite{Kasen2010magnetar,Dexter2013}. For a black hole central engine, on the other hand, instabilities in the accretion flow might produce strong light-curve variability, as seen in active galactic nuclei (AGN)\cite{Geha2003}. In this case, the light curve is expected to eventually settle onto a $t^{-5/3}$ decline rate\cite{Michel1988} after the last instability. Such a decline rate is indeed observed for {\hls} starting around day $450$ (Extended Data Fig. \ref{EDfig:addedphot}), supporting a black hole power source.

{\noindent}We conclude that {\hls} does not show the expected signatures of magnetar power (using available analytical models), but might be consistent with black hole accretion power.

\subsection{No signs of asymmetry in {\hls}.}
\label{sec:polarimetry}

A possible explanation for the high luminosities and apparent emitted energy of {\hls}, as well as the discrepancy between its line-forming vs. blackbody radii, is strong asymmetry in the explosion. Such asymmetry would be indicated by a polarization signal.

{\noindent}We observed {\hls} with the Andalucia Faint Object Spectrograph and Camera (ALFOSC) mounted on the 2.5-m Nordic Optical Telescope (NOT) in polarimetric mode on 2015 Nov.~03 in the $R$-band, and on Dec.~15 in $V$-band (we also obtained observations on 2015~Oct.~28 and Nov.~14 but we discard them because of very poor observing conditions). We used a 1/2-wave plate in the FAPOL unit and a calcite plate mounted in the aperture wheel, and observed at four different retarder angles ($0^\circ$, $22.5^\circ$, $45^\circ$, $67.5^\circ$). The data were reduced in a standard manner, using bias frames and flat fields without the polarization units in the light path. The field of view contains one bright star that can be used for calibration and for determining the interstellar polarisation (ISP) in the Galaxy. The low Galactic extinction toward {\hls} implies an expected ISP value of $< 0.13\%$\cite{Serkowski1975}. 

{\noindent}To measure the fluxes we performed aperture photometry, and to compute the polarisation we followed standard procedures\cite{Patat2006}. For our epoch with the best signal-to-noise (2015~Nov.~03), we measure $P = 0.40\pm0.27 \%$ for {\hls} and $P = 0.17\pm0.09 \%$ for the comparison star, in agreement with the ISP prediction. These results suggest that {\hls} is close to spherically symmetric, similar to what is observed for Type~II-P supernovae during their plateau phase\cite{Leonard2001}. The 2015~Dec.~15 epoch yields a lower precision ($P = 1.1\pm0.7 \%$ for {\hls} and $P = 0.80\pm0.23\%$ for the comparison star), but is still consistent with very low asphericity.

\subsection{Expansion velocities of {\hls}.}
\label{sec:velocities}

In a supernova, the ejecta are in homologous expansion --- that is, the radius of the ejecta at time $t$ evolves as $r=vt$, with faster material at larger radii. Even for perfectly mixed ejecta, at any given time, spectral lines of different elements form in different regions. Specifically, the Fe lines are formed at smaller radii than the H lines and therefore display a lower velocity. This is also the case in {\hls}. As time passes and the ejecta expand and recombine, the line-forming region of each element moves inward in mass to a region where the outflow is slower. This is why, normally, the velocity of all lines is observed to decrease with time. Thus, following the line velocity over a wide range of time (and hence mass coordinates) provides a ``scan'' of the velocity profile over a large range of the ejecta. Although different lines are formed in different regions, all line-forming regions scan the velocity profile of the same ejecta. Therefore, if there is a significant velocity gradient in the ejecta, we expect to see both a significant velocity difference between the Fe and H lines as well as significant evolution in the velocity of each line as the material expands. These two features are seen clearly in the typical case of {\snem} (Extended Data Fig. \ref{EDfig:vgradient}). However, this is not the case in {\hls}. On the one hand, there is a significant difference between the H and Fe line velocities, indicating a large velocity gradient in the ejecta. On the other hand, the velocity of each line shows almost no evolution in time between days 100 and 600 after discovery. If the line-forming material were ejected at the time of discovery, then this time span corresponds to a change by a factor of $\approx6$ in radius. In this case, the lack of observed velocity evolution indicates a very shallow velocity gradient in the ejecta, which is inconsistent with the large velocity difference between the lines. However, if the ejection of the line-forming material took place before discovery, then the relative change in radius during the observations is small, indicating that the position of the line-forming region does not change much, potentially solving the apparent contradiction.

\subsection{The line-forming region of {\hls}.}
\label{sec:lineforming}

The nearly constant line velocities measured in {\hls} suggest that the lines form in a massive shell, perhaps ejected prior to the explosion. Here we estimate the mass and energetics required for such a shell to produce the observed line features.

{\noindent}Consider a uniform shell of mass $M$ with a radius $r$ and width \dr. The number density of hydrogen atoms in the shell is
\begin{equation}
n_{\rm H} = \frac{Y_H M}{\mu m_p 4 \pi r^2 \dr },
\end{equation}
where $Y_H \approx 0.9$ is the number fraction of hydrogen and $\mu \approx 1.34$ is the mean atomic mass for solar gas ($m_p$ is the proton mass). 
In a rapidly expanding, homologous outflow ejected at a time \tej, the strength of a spectral line is characterized by the Sobolev optical depth approximation
\begin{equation}
\taus = \frac{ \pi e^2}{m_e c} n_l f \tej \lambda_0,
\end{equation}
where $n_l$ is the number density of atoms in the lower level, $f$ is the line oscillation strength, \tej\ is the time since explosion, and $\lambda_0$ is the line rest wavelength. For a line to produce a noticeable absorption component in the spectra, it must have $\taus \gtrsim 1$.  

{\noindent}To estimate the populations in the lower level of the line transition (for the Balmer series this is the $n=2$ level), we apply the nebular approximation\cite{Abbott1985}, which assumes the mean intensity of the radiation field at a radius above a nearly blackbody photosphere is $J_\nu(r) = W(r) B_\nu(T_{\rm bb})$, where $B_\nu$ is the Planck function, $T_{\rm bb}$ is the temperature of the photosphere, and $W(r)$ is the geometrical dilution factor of the radiation field,
\begin{equation}
W(r) = \frac{1}{2} \left[ 1 - \sqrt{1 - r_p^2/r^2} \right] \approx \frac{r_p^2}{4 r^2}. 
\end{equation}
Here, $r_p$  is the photospheric radius and the last expression assumes $r \gg r_p$.
For a two-level atom subject to this radiation field, the number density in the $n = 2$ excited state is
\begin{equation}
n_2 \approx n_1 W \frac{g_2}{g_1} e^{-\Delta E_{1,2}/k T}, 
\end{equation}
where ($n_1$,$n_2$) and ($g_1$,$g_2$) are respectively the number density and statistical weights of the $n=1$ and $n=2$ levels, and $\Delta E_{1,2}$ is the energy difference between the levels.  

{\noindent}Since essentially all of the hydrogen in the shell will be neutral and in the ground state, $n_1 \approx n_{\rm H}$. The Sobolev optical depth is then
\begin{equation}\label{eq:tau}
\tau_{\rm H\alpha} \approx \left[ \frac{ \pi e^2}{m_e c}   f  \lambda_0 \tej \right]
\frac{Y_H M}{\mu m_p} \frac{r_p^2} {16 \pi r^4 \dr} \frac{g_2}{g_1}   e^{-\Delta E_{1,2}/k T}.
\end{equation}
Using $g_1 = 2$, $g_2 = 8$, $\Delta E_{1,2} = 10.2\,$eV, $\lambda_0 = 6563\,\Angstrom$ (for the H$\alpha$ transition), and $f = 0.64$, and taking $T = 6500$\,K, $\dr  = \Delta v \tej$, and $r =v \tej$ gives
\begin{equation}
\tau_{\rm H\alpha} \approx 0.96 \left[ \frac{M}{45\,\Msun} \right] 
\left[ \frac{600\,{\rm days}}{\tej} \right]^4 
\left[ \frac{r_p}{1.5\times 10^{15}\,{\rm cm}} \right]^2
\left[ \frac{6000\,{\rm km\,s}^{-1}}{v} \right]^{4}
\left[ \frac{1000\,{\rm km\,s}^{-1}}{\Delta v} \right]
\end{equation}
(where $\Msun$ is the solar mass). Though approximate, this argument demonstrates that a shell with a mass of order a few tens of solar masses is likely required for producing Balmer absorption lines throughout the $\approx 600$-day duration of the {\hls} light curve. The corresponding kinetic energy of the outburst is $\sim10^{52}$\,erg. In the case that the shell was ejected before the first {\hls} observations, the mass and energy required would increase. However, the mass required to associate the line-forming region with the 1954 eruption would be $\sim10^7\,\Msun$, and hence not reasonable, implying that the line-forming region was ejected in a separate, more recent, eruption.

{\noindent}For comparison, the electron-scattering optical depth of the shell is
\begin{equation}
\tau_{\rm es} =  n_{\rm H} x_{\rm H\,II} \sigma_T \dr \approx 0.77 x_{\rm H\,II} \left[ \frac{M}{45\,\Msun} \right] \left[ \frac{600\,{\rm days}}{\tej} \right]^2 
\left[ \frac{6000~{\rm km\,s}^{-1}}{v} \right]^{4},
\end{equation}
where $\sigma_T$ is the Thomson cross-section and $x_{\rm H\,II}$ is the fraction of ionized hydrogen. The shell will be largely neutral ($x_{\rm H\,II} \ll 1$), because the region where the radiation field is sufficient to ionize hydrogen occurs at the photosphere ($r_p$) where the recombination front forms. The shell radius is much larger than $r_p$, so the radiation field is strongly diluted. Thus, while the shell can form line features, it will be optically thin in the continuum and allow most of the pseudo-blackbody continuum from the photosphere to pass through.

{\noindent}The velocity of 6000\,km\,s$^{-1}$ seen for H$\alpha$ at day 600 after discovery, is also seen for H$\beta$ at day 200 after discovery. If we calculate the optical depth (Eq. \ref{eq:tau}) for H$\beta$ (plugging in the parameters for day $200+t_0$, where $t_0$ is the offset between the ejection of the shell and discovery) and equate it to that of H$\alpha$ at day $600+t_0$, then we can solve for the ejection time $t_0$, assuming the optical depths for H$\alpha$ and H$\beta$ were the same when each was observed at 6000\,km\,s$^{-1}$, and that the entire shell was ejected simultaneously. Using $\lambda_0 = 4861\,\Angstrom$ and $f = 0.12$ for the H$\beta$ transition, we find $t_0\approx$100--200 days (the main source of error is the uncertainty in the precise temperature difference between the two epochs), meaning that the line-forming shell was ejected 100--200 days before discovery. We have deep nondetection limits for part of this epoch (Extended Data Fig. \ref{EDfig:limits}), suggesting that the ejection of the shell could have been a low-luminosity event. This estimation of the ejection time, however, relies on many simplifying assumptions, so it should be considered only as an approximation.

\subsection{A historical outburst at the position of {\hls}.}
\label{sec:poss}

The Palomar Observatory Sky Survey (POSS)\cite{Minkowski1963} observed the field of {\hls} on 1954 Feb. 23 in the blue and red filters. POSS-II\cite{Reid1991} then re-observed the field on 1993 Jan. 2 in the blue filter and on 1995 Mar. 30 in the red filter. We obtained these images through the STScI Digitized Sky Survey and we find a source at the position of {\hls} in the blue image from POSS that is not present in the blue image from POSS-II (Extended Data Fig. \ref{EDfig:poss}). We do not see this source in either of the red images, but they are not as deep as the blue images (the limiting magnitude is roughly 20 for the red images compared to 21.1 for the blue images)\cite{Minkowski1963}.

{\noindent}We register the POSS blue image to the POSS-II blue image using the IRAF task {\tt wregister}. We then use the {\tt apphot} package in PyRAF, with a 3-pixel aperture, to measure the flux in six stars in the field near the position of {\hls} to determine a zero-point offset for the two images. We find an offset of $0.132\pm0.050$ mag. We then perform the same measurement around the nucleus of the host galaxy of {\hls} and find an offset of $0.141$ mag, consistent with the zero-point offset. Next we perform the same aperture photometry measurement at the position of {\hls} in both images. We find a magnitude difference of $0.31\pm0.14$ over the host-galaxy level confirming the presence of an outburst in the 1954 image at the position of {\hls} at a $2.2\sigma$ confidence level.  Owing to the nonlinear nature of the photographic plates used in the two POSS surveys, as well as differences between the filters\cite{Reid1991}, we cannot perform meaningful image subtraction between the POSS epochs to obtain more accurate photometric measurements. We consider this confidence level to be a conservative estimate, the outburst can be seen clearly by eye in the images (Extended Data Fig. \ref{EDfig:poss}).

{\noindent}We calibrate the six stars used for the zero-point comparison to SDSS $u$-band plus $g$-band fluxes (the POSS blue filter roughly covers the SDSS $u$ and $g$ bands)\cite{Minkowski1963} and find that the magnitude of the 1954 outburst (after removing the host-galaxy contribution) is $20.4\pm0.1\,\textrm{(stat)}\pm0.8\,\textrm{(sys)}$. The first error is statistical and due to photometric measurement uncertainties, while the second error is systematic and caused by the calibration to SDSS (the large error value is likely produced by filter and detector differences between POSS and SDSS).

{\noindent}This corresponds to an absolute magnitude for the outburst of $\approx-15.6$ at the luminosity distance of {\hls} (this is only a lower limit on the peak luminosity of the eruption, as we have only one epoch of observations). Such an eruption may be produced by the pulsational pair instability\cite{Barkat1967,Heger2002,Woosley2007,Woosley2017}. Eruptions of similar luminosity (though likely caused by different instabilities) are inferred to be common in Type IIn supernova progenitors in the last year prior to explosion\cite{Ofek2014}. Spectra and broad-band colors are available for three such possible outbursts --- a precursor to PTF10bjb\cite{Ofek2014}, PTF13efv (a precursor to SNHunt275)\cite{Ofek2016} and the first 2012 outburst of SN\,2009ip\cite{Fraser2013} --- all of which display rather flat continuum emission, consistent with the limited color information we have for the 1954 outburst of {\hls} (i.e. the red nondetection limit being $\approx0.4$ magnitudes brighter than the blue detection).

{\noindent}Given the host galaxy size of $\sim$10--100 times the centroiding error of the outburst, and a typical supernova rate of $\sim$1/100 per galaxy per year, there is a few percent probability that the detected outburst is an unrelated supernova that happened to occur at the position of {\hls}. 

\subsection{The rate of {\hls}-like events.}
\label{sec:rate}

On 2014 Nov. 18, {\hls} was independently discovered by the Catalina Real-Time Transient Survey\cite{Drake2009} as CSS141118:092034+504148, and more recently the event was reported to the Transient Name Server as AT\,2016bse and Gaia16aog. The fact that it was discovered multiple times, but dismissed as a run-of-the-mill Type~II-P supernova, suggests that similar events may have been missed in the past. We ourselves would not have noticed the unique properties of {\hls} had the iPTF survey scheduler not automatically continued to monitor the position of {\hls}. In addition, if {\hls}-like events are limited to low-mass galaxies, then targeted transient surveys would have missed them completely. 

To our knowledge, {\hls} is the only supernova ever discovered to show such long-lived, slowly evolving II-P-like emission. The PTF and iPTF surveys discovered $631$ Type~II supernovae, indicating that {\hls}-like events could be $\sim10^{-3}$--$10^{-2}$ of the Type~II supernova rate. Since luminous, long-lived varying events could be easier to detect in transient surveys compared to normal supernovae, the true volumetric rate of {\hls}-like events could be much lower. On the other hand, we cannot rule out whether such events were discovered in the past but dismissed as normal Type~II-P supernovae after one spectrum with no subsequent follow-up observations or as possible AGN owing to the light curve behavior. It is therefore not possible to calculate a precise rate for {\hls}-like events based on this single discovery, but whatever the explosion channel, it is likely to be rare. Even so, the Large Synoptic Survey Telescope could find hundreds of {\hls}-like events in its decade-long survey of the transient sky (more so if {\hls}-like events are more common in the early Universe, as is indicated by the possible low-metallicity environment of {\hls}). 

\subsection{The host galaxy of {\hls}.}
\label{sec:host}

We obtained a spectrum of the host galaxy of {\hls} on 2015 Dec. 11 with the Low Resolution Imaging Spectrometer (LRIS)\cite{Oke1995} mounted on the Keck-I 10-m telescope. The spectrum was reduced using standard techniques optimized for Keck+LRIS by the {\tt CarPy} package in PyRAF, and flux calibrated with spectrophotometric standard stars obtained on the night of our observations in the same instrument configuration. 
The host-galaxy spectrum, which is available for download via WISeREP\cite{Yaron2012}, shows clear detections of H$\alpha$, H$\beta$, [O II] $\lambda3727$ and [O III] $\lambda\lambda4958$,$5007$, which we use to determine a redshift of 0.0344. A faint detection of [N II] $\lambda6583$ is also possible, but the feature is difficult to confirm because the continuum is contaminated by broad H$\alpha$ emission from the nearby supernova. All of the lines are weak (equivalent width $<20\Angstrom$) and no other lines are significantly detected. We extracted the fluxes of all lines by fitting Gaussians to their profiles (Extended Data Table \ref{EDtab:host_lines}), and calculated the metallicity by fitting\cite{Bianco2016} the line-strength ratios using several different diagnostics and calibrations (Extended Data Table \ref{EDtab:host_metallicity}). We find a range of metallicity estimates of $12+{\log}\,\textrm{(O/H)}=$8.3--8.6, corresponding to $\approx$0.4--0.9\,$\Zsun$ (where $\Zsun$ is the solar metallicity)\cite{Asplund2009}. A low metallicity could help explain how the progenitor of {\hls} retained a very massive hydrogen envelope. Future, more direct environment studies will be able to better probe the metallicity at the explosion site.

{\noindent}We fit the SDSS $ugriz$ photometry of the host galaxy\cite{alam2015} with standard SED fitting techniques\cite{Perley2013} using the BC03\cite{Bruzual2003} stellar population synthesis models. Assuming a metallicity of $0.5\,\Zsun$, the best-fit total stellar mass is $\left(3.2\pm0.5\right)\times10^8$\,$\Msun$, similar to that of the Small Magellanic Cloud.


\end{methods}


\begin{das}

The photometric data that support the findings of this study are available in the Open Supernova Catalog\cite{Guillochon17}, https://sne.space/sne/iPTF14hls/. The spectroscopic data that support the findings of this study are available on the Weizmann Interactive Supernova data REPository (WISeREP)\cite{Yaron2012}, https://wiserep.weizmann.ac.il/, and on the Open Supernova Catalog. Source data for figures \ref{fig:photometry}, \ref{fig:velocities} and \ref{fig:radius}, and for Extended Data Figures \ref{EDfig:addedphot}, \ref{EDfig:limits} and \ref{EDfig:superfits} are provided with the paper. 

\end{das}

%

\newpage

\begin{EDfigure}
\centering
\includegraphics[width=15cm]{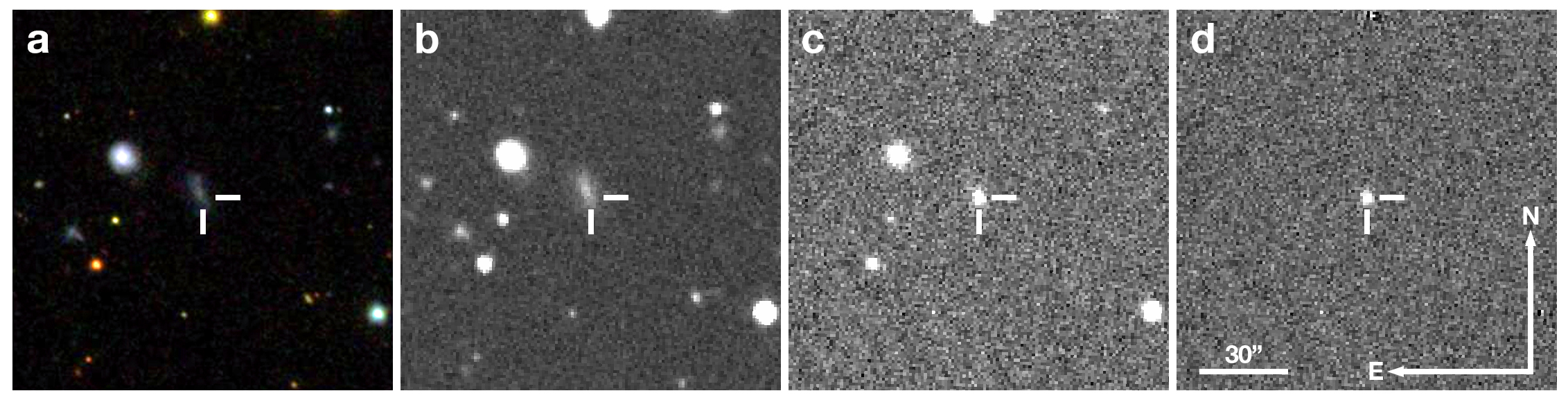}
\caption{{\bf The discovery and environment of {\hls}.} (\textbf{a}) SDSS image centered at the position of {\hls}. (\textbf{b}) Palomar 48-inch deep coadded pre-discovery reference image. (\textbf{c}) Palomar 48-inch discovery image of {\hls}. (\textbf{d}) The result of subtracting the reference image from the discovery image. The position of {\hls} is indicated by tick marks in each image.
\label{EDfig:discovery}}
\end{EDfigure}

\newpage

\begin{EDfigure}
\centering
\includegraphics[width=15cm]{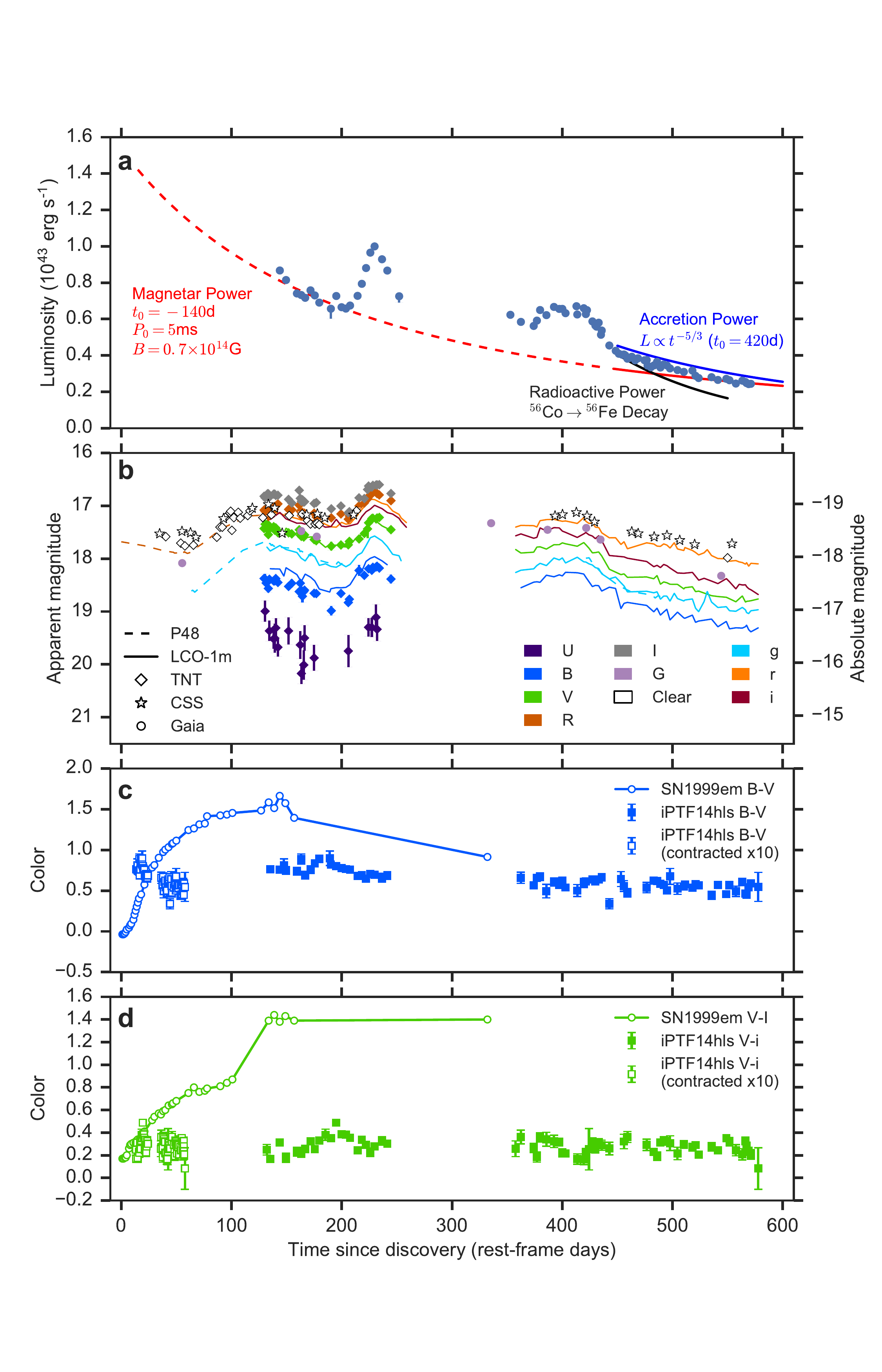}
\caption{{\bf Additional photometry of {\hls}.} The bolometric light curve of {\hls} (\textbf{a}) deduced from the blackbody fits shows a late-time decline rate which is slower than the radioactive decay of $^{56}$Co (black), but consistent with both delayed accretion power (blue; $t_0$ is the onset of accretion at the last peak which could represent a final fallback event) and magnetar spindown power (red; $t_0$ is the formation time of the magnetar, $P_0$ is the initial spin period and $B$ is the magnetic field in this simple analytic model). The magnetar model, however, is not consistent with the luminosity during the first 100 days, as implied by the P48, CSS and Gaia observations (\textbf{b}), unless the early-time magnetar emission is significantly adiabatically degraded. TNT photometry of {\hls} and publicly available CSS photometry (retrieved from the CSS website) and Gaia photometry (retrieved from the Gaia Alerts website) not presented in Figure \ref{fig:photometry} are shown in panel (\textbf{b}). Data from the P48 (dashed lines) and the LCO 1-m telescope (solid lines) presented in Figure \ref{fig:photometry} are shown for comparison. Photometric points from the same day, instrument, and filter are averaged for clarity. The {\it B-V} (\textbf{c}) and {\it V-I/i}  (\textbf{d}) color evolution of {\hls} from the LCO 1-m data (filled squares) differs from that of the normal Type~II-P {\snem} (empty circles)\cite{Leonard2002}, even when contracting the {\hls} data by a factor of 10 in time (empty squares) to compensate for the slow evolution observed in its spectra compared to that of normal II-P supernovae. All error bars, when available, denote $1\sigma$ uncertainties. 
\label{EDfig:addedphot}}
\end{EDfigure}

\newpage

\begin{EDfigure}
\centering
\includegraphics[width=15cm]{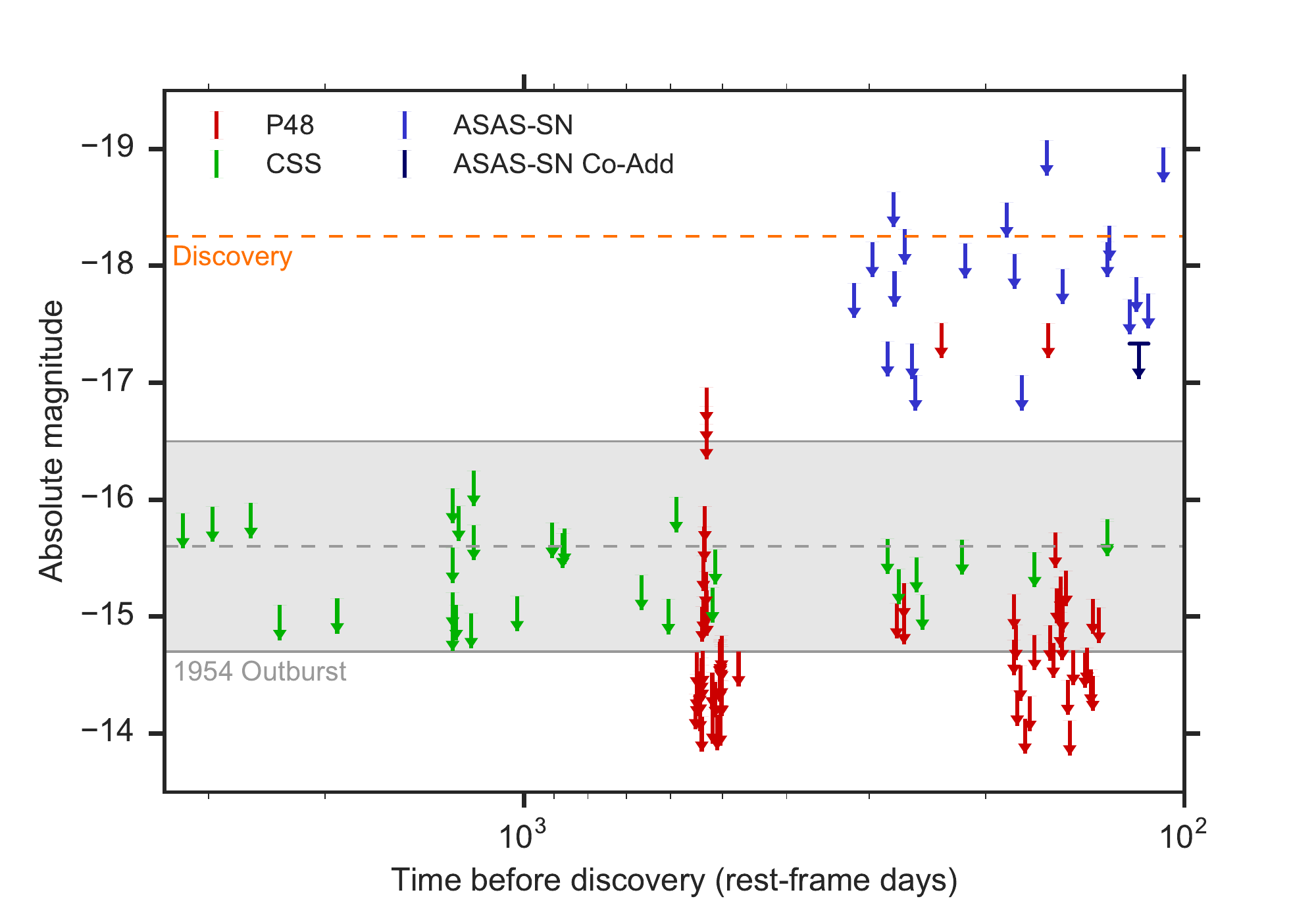}
\caption{{\bf Pre-explosion nondetection limits for {\hls}.} Data from P48 ($R$ band, $3\sigma$ nondetections), CSS (unfiltered, obtained via the CSS website), and ASAS-SN ($V$-band, $3\sigma$ nondetections --- the dark-blue arrow is a deep coadd of the three images taken during the time range denoted by the horizontal line in the marker) are shown. The dashed line indicates the discovery magnitude and the shaded region shows the 1954 outburst magnitude and its uncertainty.
\label{EDfig:limits}}
\end{EDfigure}

\newpage

\begin{EDfigure}
\centering
\includegraphics[width=15cm]{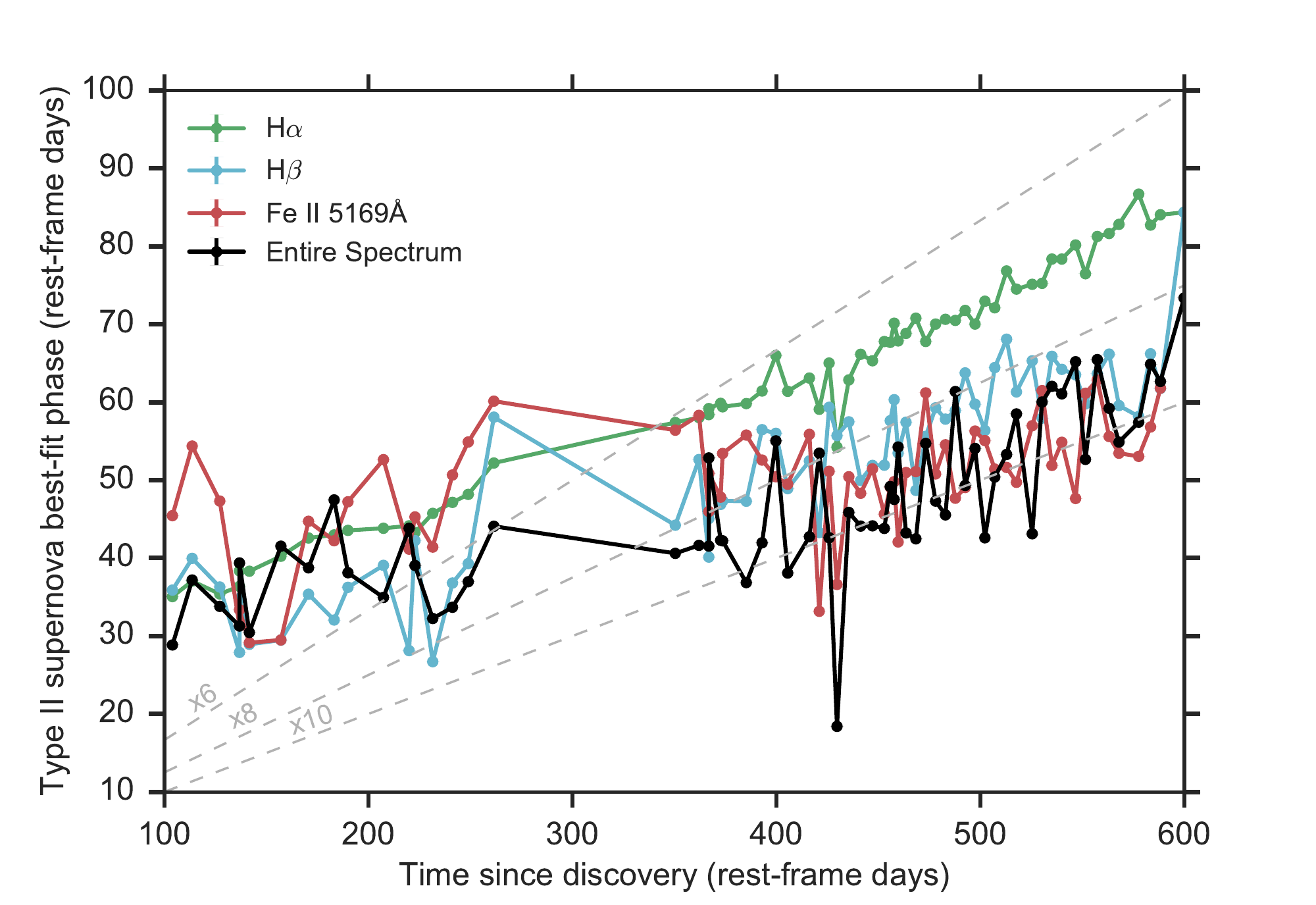}
\caption{{\bf The slow spectral evolution of {\hls} compared to normal supernovae.} Weighted average best-fit phase of {\hls} spectra from Superfit\cite{Howell2005}, compared to the true spectral phase are shown, when fitting the entire spectrum (black) or only certain line regions as noted. The dashed lines denote constant ratios between the observed and best-fit phases (assuming the explosion happened at discovery). The spectra of {\hls} evolve a factor of $\approx$6--10 slower than those of other Type~II supernovae.
\label{EDfig:superfits}}
\end{EDfigure}

\newpage

\begin{EDfigure}
\centering
\includegraphics[width=15cm]{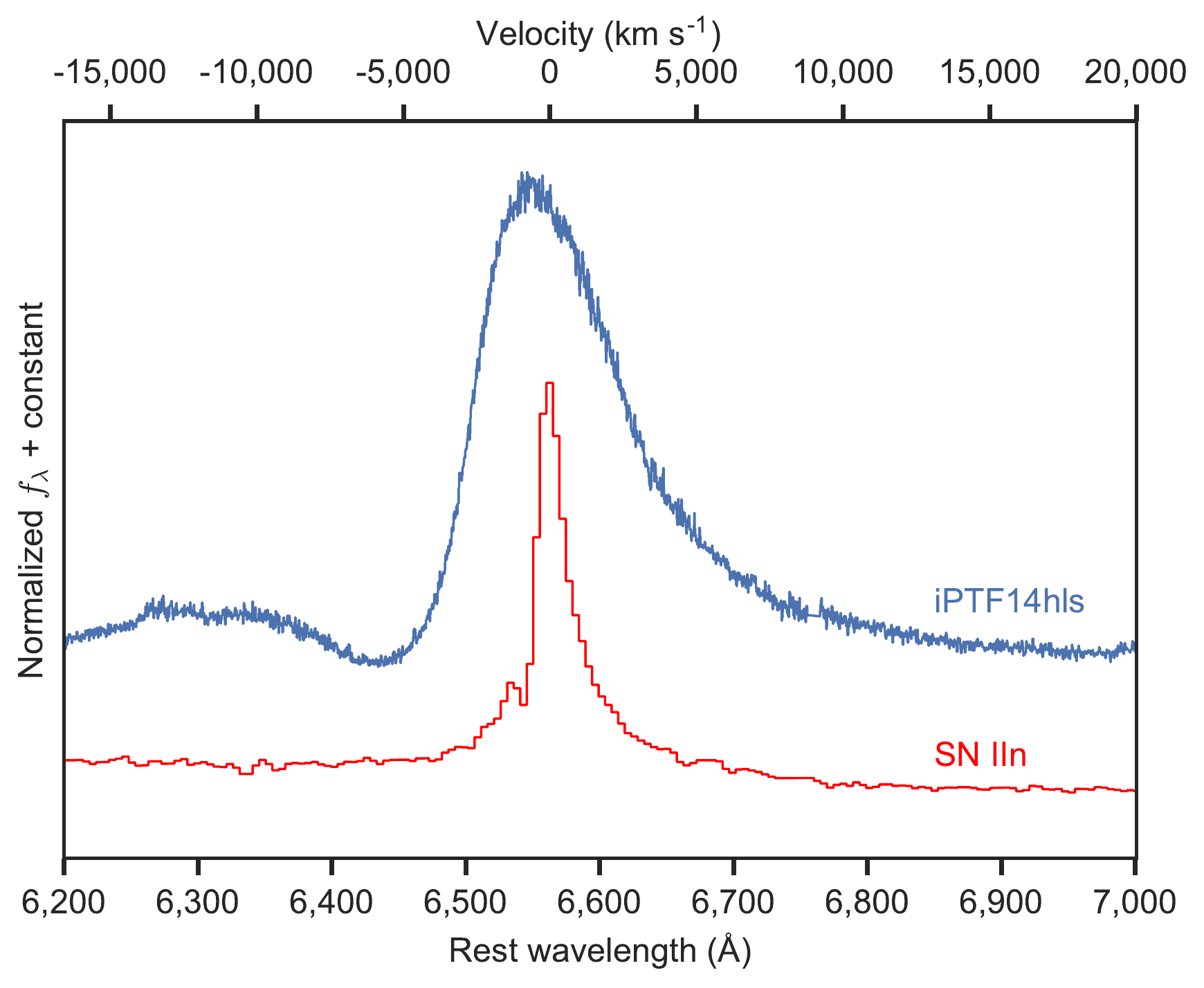}
\caption{{\bf A lack of spectral interaction signatures in {\hls}.} The H$\alpha$ region in our highest-resolution spectrum of {\hls} taken on 2016 June 4 using DEIMOS on Keck II (blue), expressed in terms of normalized flux density as a function of rest-frame wavelength (bottom axis), compared to the interaction-powered Type IIn SN\,2005cl\cite{Kiewe2012} (red). The top axis is the corresponding velocity of H$\alpha$. {\hls} shows no signs of the narrow emission or narrow P-Cygni features seen in interacting supernovae.
\label{EDfig:deimos}}
\end{EDfigure}

\newpage

\begin{EDfigure}
\centering
\includegraphics[width=15cm]{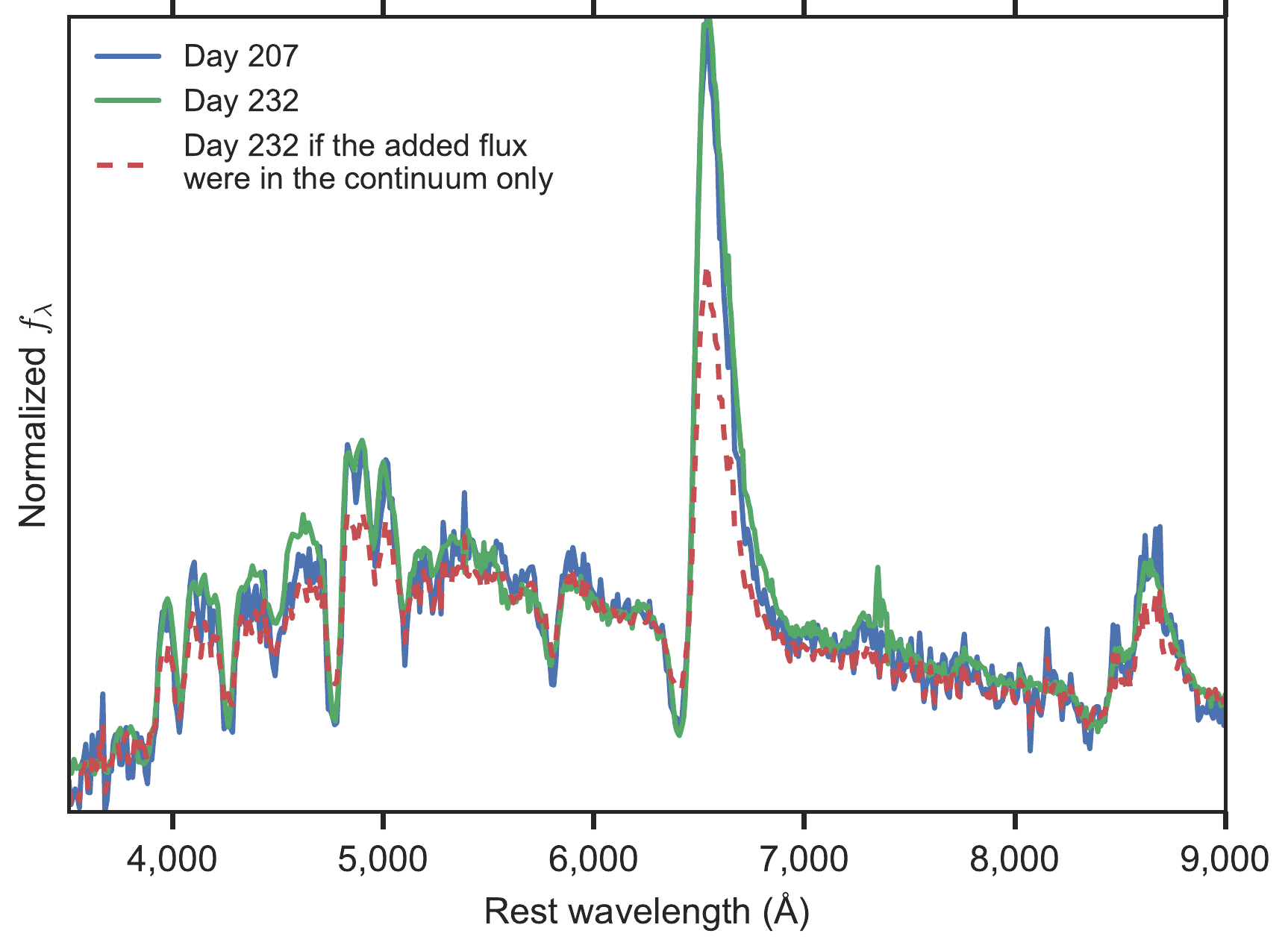}
\caption{{\bf The nature of the increased flux during the brightest peak of {\hls}.} Spectra of {\hls} expressed in terms of normalized flux density as a function of rest-frame wavelength taken on rest-frame days 207 (right before the rise to the brightest peak in the light curve) and 232 (at the brightest peak in the light curve) after discovery (solid lines) are shown. The similarity of the spectra indicates that the increase of $\approx50\%$ in luminosity observed in the light curve between the two epochs is equal at all wavelengths. If the increase were only due to the continuum flux, then the line emission on day 232 would have been diluted by the continuum (as simulated by the dashed line).
\label{EDfig:addedflux}}
\end{EDfigure}

\newpage

\begin{EDfigure}
\centering
\includegraphics[width=15cm]{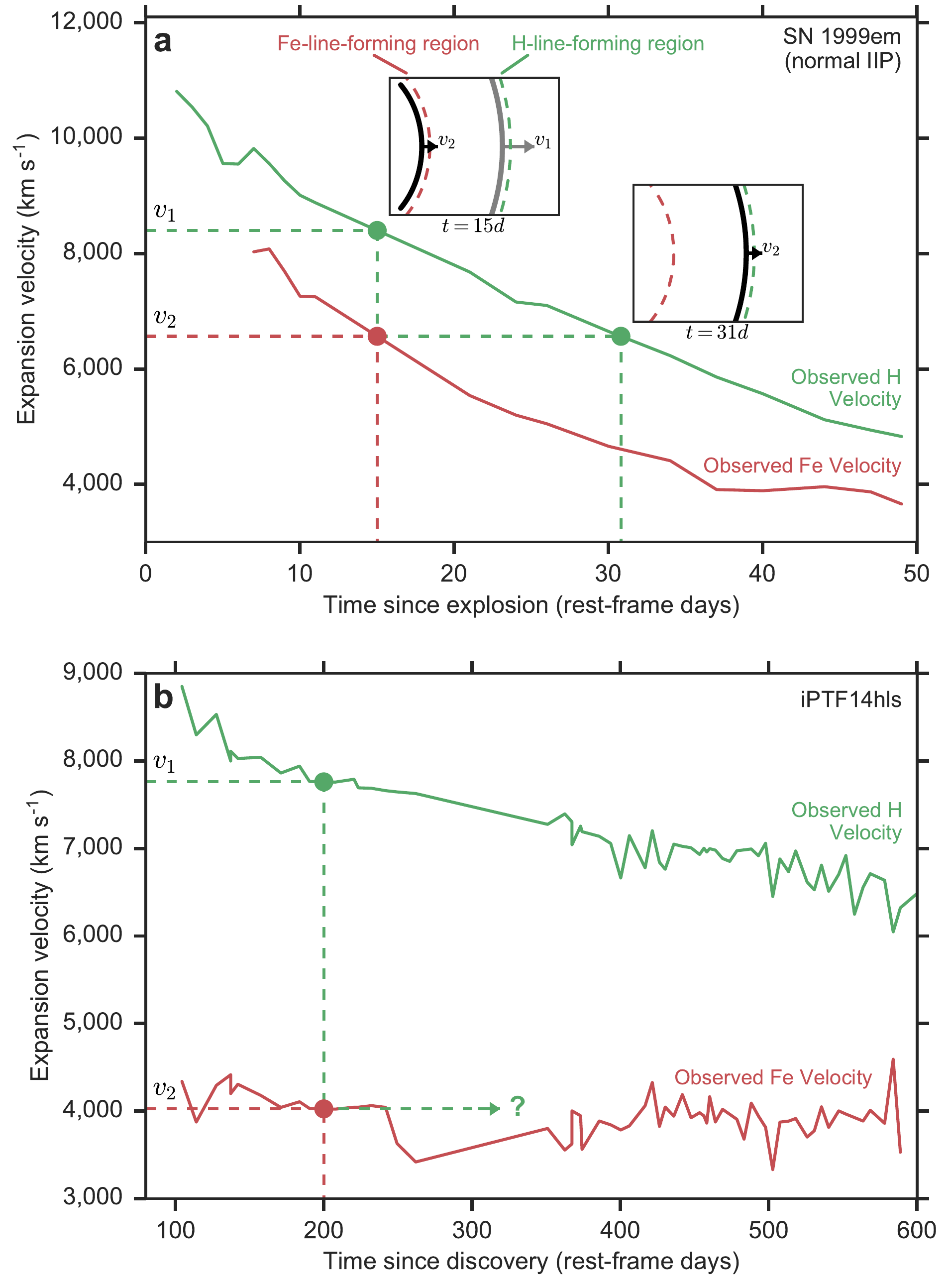}
\caption{{\bf The perplexing velocity evolution of {\hls}.} Evolution of the measured velocity gradient in the normal Type~II-P {\snem}\cite{Leonard2002} (\textbf{a}) and in {\hls} (\textbf{b}) are shown. At a given time, the H-line-forming region is at material expanding with velocity $v_1$, while the Fe-line-forming region is at material expanding with lower velocity $v_2$ (top inset in panel \textbf{a}). For {\snem}, the H-line-forming region soon reaches the material expanding at velocity $v_2$ as it moves inward in mass (bottom inset in panel \textbf{a}) and $v_2$ is measured in the H lines. For {\hls}, in contrast, the H-line-forming region does not reach the material expanding at $v_2$ even after the time since discovery increases by a factor of $6$. If the material were ejected at the time of discovery, this would indicate an increase in the radius of the line-forming regions by a factor of $\approx6$, which is unlikely given the observed velocity gradient between the H and Fe lines. If the material were ejected before discovery, on the other hand, the relative expansion in radius would be much smaller, thus offering one possible explanation for the constant velocity gradient observed in {\hls}.
\label{EDfig:vgradient}}
\end{EDfigure}

\newpage

\begin{EDfigure}
\centering
\includegraphics[width=15cm]{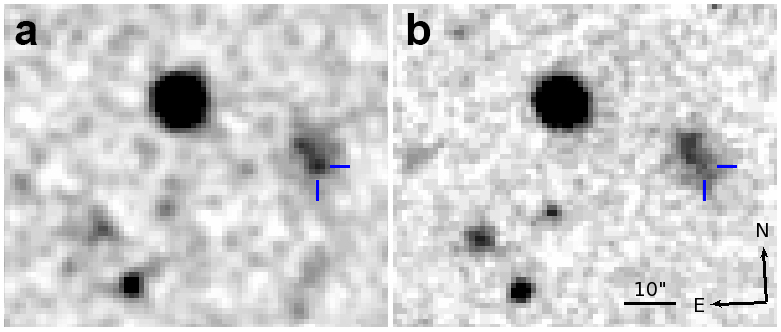}
\caption{{\bf A historic eruption at the position of {\hls}.} Blue-filter images of the position of {\hls} (marked by blue ticks) from 1954 Feb. 23 (POSS; \textbf{a}) and 1993 Jan. 2 (POSS-II; \textbf{b}) are shown. A source is visible at the position of {\hls} in the 1954 image, which is not there in the 1993 image. Using aperture photometry, we find that the 1954 source is $0.31\pm0.14$ mag brighter than the underlying host galaxy at that position, corresponding to a rough outburst magnitude of $\approx-15.6$ at the luminosity distance of {\hls}, after removing the host-galaxy contribution and calibrating the field to the SDSS $u$+$g$-bands.
\label{EDfig:poss}}
\end{EDfigure}

\newpage

\begin{EDtable}
\begin{center}
\begin{tabular}{lll}
\hline\hline
Line & Flux & Flux Error \\
\hline
$[$O~II$]$ 3727\,$\Angstrom$   &  $2.050\times10^{-16}$   &  $1.152\times10^{-17}$ \\
H$\beta$                             &  $5.666\times10^{-17}$  &  $6.349\times10^{-18}$ \\
$[$O~III$]$ 4958\,$\Angstrom$  &  $1.742\times10^{-17}$  &  $6.130\times10^{-18}$ \\
$[$O~III$]$ 5007\,$\Angstrom$  &  $1.003\times10^{-16}$  &  $6.171\times10^{-18}$ \\
H$\alpha$                           &  $1.539\times10^{-16}$  &  $4.089\times10^{-18}$ \\
$[$N~II$]$ 6583\,$\Angstrom$   &  $1.361\times10^{-17}$   &  $4.095\times10^{-18}$ \\
\hline\hline
\end{tabular}
\end{center}
\caption{{\bf {\hls} host-galaxy line fluxes.} Data are shown in erg\,s$^{-1}$\,cm$^{-2}$. Errors denote $1\sigma$ uncertainties.
\label{EDtab:host_lines}}
\end{EDtable}

\newpage

\begin{EDtable}
\begin{center}
\begin{tabular}{llll}
\hline\hline
Diagnostic & Metallicity & Lower Error & Upper Error \\
\hline
N06-N2\cite{N06}           &  $8.339$  &  $-0.126$  &  $+0.098$ \\
N06-R23\cite{N06}         &  $8.633$  &  $-0.166$  &  $+0.071$ \\
D02\cite{D02}                 &  $8.334$  &  $-0.166$  &  $+0.139$ \\
PP04-N2Ha\cite{PP04}  &  $8.250$  &  $-0.059$  &  $+0.044$ \\
PP04-O3N2\cite{PP04}  &  $8.309$  &  $-0.051$  &  $+0.037$ \\
M08-N2Ha\cite{M08}      &  $8.458$  &  $-0.116$  &  $+0.076$ \\
M13-O3N2\cite{M13}     &  $8.252$  &  $-0.035$  &  $+0.025$ \\
M13-N2\cite{M13}          &  $8.249$  &  $-0.078$  &  $+0.060$ \\
KK04-N2Ha\cite{KK04}  &  $8.490$  &  $-0.127$  &  $+0.080$ \\
KD02comb\cite{KD02}   &  $8.386$  &  $-0.130$  &  $+0.055$ \\

\hline\hline
\end{tabular}
\end{center}
\caption{{\bf {\hls} host-galaxy metallicity values.} Data are shown as $12+{\log}\,\textrm{(O/H)}$ values under different diagnostics and calibrations. Error ranges denote $1\sigma$ uncertainties.
\label{EDtab:host_metallicity}}
\end{EDtable}


\clearpage

\begin{si}

We are grateful to C. Harris, D. C. Leonard, D. Poznanski, N. Smith and S. E. Woosley for comments and discussion, and to T. Pursimo for assistance with the polarimetry measurements. 

Support for I.~Arcavi was provided by the National Aeronautics and Space Administration (NASA) through the Einstein Fellowship Program, grant PF6-170148. D.A.H., G.H., and C.M. are supported by the U.S. National Science Foundation (NSF) grant AST-1313484. This research is funded in part by the Gordon and Betty Moore Foundation through Grant GBMF5076 to L.B. and D.K. and by the NSF under grant PHY-1125915. A.G.-Y. is supported by the EU via ERC grants No. 307260 and 725161, the Quantum Universe I-Core program by the Israeli Committee for Planning and Budgeting, and the ISF; a Binational Science Foundation ``Transformative Science'' grant; and by a Kimmel award. J.S. gratefully acknowledges support from the Knut and Alice Wallenberg Foundation. P.E.N. acknowledges support from the DOE through DE-FOA-0001088, Analytical Modeling for Extreme-Scale Computing Environments. K.M. acknowledges funding from the Hintze Trust. K.J.S. is supported by the NASA Astrophysics Theory Program (grants NNX15AB16G and NNX17AG28G). X.W., T.Z., and J.Z. are supported by  Major State Basic Research Development Program (2013CB834903), the National Natural Science Foundation of China (NSFC grants: 11325313, 11403096 and 11633002), and the Strategic Priority Research Program of Emergence of Cosmological Structures of the Chinese Academy of Sciences (grant No. XDB09000000). B.S. is supported by NASA through Hubble Fellowship grant HF-51348.001 awarded by the Space Telescope Science Institute, which is operated by the Association of Universities for Research in Astronomy, Inc., for NASA, under contract NAS 5-26555. T.W.-S.H. is supported by the U.S. Department of Energy Computational Science Graduate Fellowship, grant number DEFG02-97ER25308. A.V.F.'s supernova group at U.C. Berkeley has been supported by Gary \& Cynthia Bengier, the Christopher R. Redlich Fund, the TABASGO Foundation, NSF grant AST-1211916, and the Miller Institute for Basic Research in Science. M.S. acknowledges support from EU/FP7-ERC grant 615929. Research support to I.~Andreoni is provided by the Australian Astronomical Observatory (AAO).

The Intermediate Palomar Transient Factory project is a scientific collaboration among the California Institute of Technology, Los Alamos National Laboratory, the University of Wisconsin (Milwaukee), the Oskar Klein Center, the Weizmann Institute of Science, the TANGO Program of the University System of Taiwan, and the Kavli Institute for the Physics and Mathematics of the Universe. LANL participation in iPTF was funded by the U.S. Department of Energy as part of the Laboratory Directed Research and Development program. Part of this research was carried out at the Jet Propulsion Laboratory, California Institute of Technology, under a contract with NASA. This paper made use of data from Las Cumbres Observatory global network of telescopes; the W.~M. Keck Observatory, which is operated as a scientific partnership among the California Institute of Technology, the University of California, and NASA (the observatory was made possible by the generous financial support of the W.~M. Keck Foundation); the Nordic Optical Telescope, operated by the Nordic Optical Telescope Scientific Association at the Observatorio del Roque de los Muchachos, La Palma, Spain, of the Instituto de Astrofisica de Canarias; ALFOSC, which is provided by the Instituto de Astrofisica de Andalucia (IAA) under a joint agreement with the University of Copenhagen and NOTSA; DOLoRes on TNG; the Discovery Channel Telescope (DCT) at Lowell Observatory. Lowell is a private, nonprofit institution dedicated to astrophysical research and public appreciation of astronomy and operates the DCT in partnership with Boston University, the University of Maryland, the University of Toledo, Northern Arizona University and Yale University. The upgrade of the DeVeny optical spectrograph has been funded by a generous grant from John and Ginger Giovale. We thank the support of the staffs at {\it Swift}, the Keck Observatory, AMI-LA, the Xinglong Observatory (part of the National Astronomical Observatories of China), and the Lijiang Observatory (part of the Yunnan Observatories of China) for assistance with the observations. The AMI is supported by the European Research Council. Development of ASAS-SN has been supported by NSF grant AST-0908816 and CCAPP at the Ohio State University. ASAS-SN is supported by NSF grant AST-1515927, the Center for Cosmology and AstroParticle Physics (CCAPP) at OSU, the Mt. Cuba Astronomical Foundation, George Skestos, and the Robert Martin Ayers Sciences Fund. ASAS-SN thanks LCO and its staff for their continued support. The Digitized Sky Surveys were produced at the Space Telescope Science Institute under U.S. Government grant NAG W-2166.  The National Geographic Society Palomar Observatory Sky Atlas (POSS-I) was made by the California Institute of Technology with grants from the National Geographic Society. The Second Palomar Observatory Sky Survey (POSS-II) was made by the California Institute of Technology with funds from the NSF, the National Geographic Society, the Sloan Foundation, the Samuel Oschin Foundation, and the Eastman Kodak Corporation. This research used resources of the National Energy Research Scientific Computing Center, a DOE Office of Science User Facility supported by the Office of Science of the U.S. Department of Energy under Contract No. DE-AC02-05CH11231.

\end{si}

\end{document}